\documentclass[preprint,showpacs,preprintnumbers,amsmath,amssymb]{revtex4}

\usepackage{graphicx}
\usepackage{dcolumn}
\usepackage{bm}

\usepackage{amsthm}

\renewcommand{\vec}[1]{\hm{#1}}
\def\i{\imath}

\def\drawgridA{%
\put(2,1){$0$}%
\put(0,5){\vector(1,0){130}}%
\put(123,1){$2\pi$}%
\put(128,7){$q$}%
\put(5,0){\vector(0,1){45}}%
\put(1,42){$A$}%
\put(5,41){\line(-1,0){1}}%
\put(2,37){$1$}%
\put( 15,5){\line(0,-1){1}}%
\put( 25,5){\line(0,-1){1}}%
\put( 35,5){\line(0,-1){1}}%
\put( 45,5){\line(0,-1){1}}%
\put( 55,5){\line(0,-1){1}}%
\put( 65,5){\line(0,-1){1}}%
\put( 75,5){\line(0,-1){1}}%
\put( 85,5){\line(0,-1){1}}%
\put( 95,5){\line(0,-1){1}}%
\put(105,5){\line(0,-1){1}}%
\put(115,5){\line(0,-1){1}}%
\put(125,5){\line(0,-1){1}}%
\put( 15,5){\dashbox{1}(0,36){}}
\put( 25,5){\dashbox{1}(0,36){}}
\put( 35,5){\dashbox{1}(0,36){}}
\put( 45,5){\dashbox{1}(0,36){}}
\put( 55,5){\dashbox{1}(0,36){}}
\put( 65,5){\dashbox{1}(0,36){}}
\put( 75,5){\dashbox{1}(0,36){}}
\put( 85,5){\dashbox{1}(0,36){}}
\put( 95,5){\dashbox{1}(0,36){}}
\put(105,5){\dashbox{1}(0,36){}}
\put(115,5){\dashbox{1}(0,36){}}
\put(125,5){\dashbox{1}(0,36){}}
}

\newtheorem{_theorem}{Theorem}

\begin{document}

\title{Stability analysis of dynamical regimes in nonlinear systems with
discrete symmetries}

\author{G.\,M.~Chechin}
 \email{chechin@phys.rsu.ru}
\author{K.\,G.~Zhukov}
\affiliation{Department of Physics, Rostov State University, Russia}

\date{\today}

\begin{abstract}
We present a theorem that allows to simplify linear stability
analysis of periodic and quasiperiodic nonlinear regimes in
$N$-particle mechanical systems (both conservative and dissipative)
with different kinds of discrete symmetry. This theorem suggests a
decomposition of the linearized system arising in the standard
stability analysis into a number of subsystems whose dimensions can
be considerably less than that of the full system. As an example of
such simplification, we discuss the stability of bushes of modes
(invariant manifolds) for the Fermi-Pasta-Ulam chains and prove
another theorem about the maximal dimension of the above mentioned
subsystems.
\end{abstract}


\maketitle

\section{\label{intro}Introduction}

Different dynamical regimes in a mechanical system with
discrete-symmetry group $G_0$ can be classified by subgroups
$G_j\subseteq G_0$ of this group \cite{DAN1,DAN2,PhysD}. Actually,
we can find an invariant manifold corresponding to each subgroup
$G_j$ and decompose it into the basis vectors of the irreducible
representations of the group $G_0$. As a result of this procedure,
we obtain a \emph{bush of modes} (see above cited papers) which can
be considered as a certain physical object in geometrical, as well
as dynamical sense. The mode structure of a given bush is fully
determined by its symmetry group $G_j$ and is independent of the
specific type of interparticle interactions in the system. In
Hamiltonian systems, bushes of modes represent dynamical objects in
which the energy of initial excitation turns out to be ``trapped''
(this is a phenomenon of energy localization in the modal space).
The number of modes belonging to a given bush (the bush dimension)
does not change in time, while amplitudes of the modes do change,
and we can find dynamical equations determining their evolution.

Being an \emph{exact} nonlinear excitation, in the considered
mechanical system, each bush possesses its own domain of
\emph{stability} depending on the value of its mode amplitudes.
Beyond the stability threshold a phenomenon similar to the
parametric resonance occurs, the bush loses its stability and
transforms into another bush of higher dimension. This process is
accompanied by spontaneous lowering of the bush symmetry:
$G_j\rightarrow \tilde{G_j}$, where $\tilde{G_j}\subset G_j$.

The concept of bushes of modes was introduced in \cite{DAN1,DAN2},
the detailed theory of these dynamical objects was developed in
\cite{PhysD}. Low-dimensional bushes in mechanical systems with
various kinds of symmetry and structures were studied in
\cite{DAN1,DAN2,PhysD,IntJ,ENOC,Octa,C60,FPU1,FPU2}. The problem of
bush stability was discussed in \cite{PhysD,Octa,FPU1,FPU2}. Two
last papers are devoted to the vibrational bushes in the
Fermi-Pasta-Ulam (FPU) chains.

Note that dynamical objects equivalent to the bushes of modes were
recently discussed for the monoatomic chains in the papers of
different authors \cite{PR,BR,Shin,AntiFPU}. Let us emphasize that
group-theoretical methods developed in our papers
\cite{DAN1,DAN2,PhysD} can be applied efficiently not only to the
monoatomic chains (as was illustrated in \cite{FPU1,FPU2}), but to
\emph{all} other physical systems with discrete symmetry groups
(see, \cite{DAN1,DAN2,PhysD,IntJ,ENOC,Octa,C60}).

In this paper, we present a theorem which can simplify essentially
the stability analysis of the bushes of modes in complex systems
with many degrees of freedom. The usefulness of this theorem is
illustrated with the example of nonlinear chains with a large number
of particles. Note that the simplification of the stability analysis
in such systems actually originate from the well-known Wigner
theorem about the block-diagonalization of the matrix commuting with
all matrices of a representation of a given symmetry group.

In Sec.\,\ref{Sec2}, we start with the simplest examples for
introducing basic concepts and ideas. In Sec.\,\ref{Sec3}, we
present a general theorem about invariance of the dynamical
equations linearized near a given bush with respect to the bush
symmetry group. In Sec.\,\ref{Sec4}, we prove a theorem which turns
out to be very useful for splitting the above mentioned linearized
dynamical equations for $N$-particle monoatomic chains. Some results
on the bush stability in the FPU chains are discussed in
Sec.\,\ref{Sec5}.

\section{Some simple examples\label{Sec2}}

\subsection{FPU-chains and their symmetry}

We consider \emph{longitudinal} vibrations of $N$-particle chains of
identical masses ($m=1$) and identical springs connecting
neighboring particles. Let $x_i(t)$ be the displacement of the
$i$-th particle ($i=1,2,...,N$) from its equilibrium position at a
given instance $t$. Dynamical equations of such mechanical system
(FPU-chain) can be written as follows:
\begin{equation}
\ddot{x}_i=f(x_{i+1}-x_i)-f(x_i-x_{i-1}).
 \label{eq1}
\end{equation}
The nonlinear force $f(x)$ depends on the deformation $x$ of the
spring as $f(x)=x+x^2$ and $f(x)=x+x^3$ for the FPU-$\alpha$ and
FPU-$\beta$ chains, respectively. We assume the periodic boundary
conditions
\begin{equation}
x_0(t)\equiv x_N(t),\qquad x_{N+1}(t)\equiv x_1(t)
 \label{eq2}
\end{equation}
to be valid. Let us also introduce the ``configuration vector''
$\vec{X}(t)$ which is the $N$-dimensional vector describing all the
displacements of the individual particles at the moment $t$:
\begin{equation}
\vec{X}(t)=\{x_1(t),x_2(t),\dots,x_N(t)\}.
 \label{eq3}
\end{equation}

In the equilibrium state, a given chain is invariant under the
action of the operator $\hat a$ which shifts the chain by the
lattice spacing $a$. This operator generates the translational group
\begin{equation}
T_N=\{\hat e,\hat a,\hat a^2,\dotsc,\hat a^{N-1}\},\quad\hat
a^N=\hat e, \label{eq4_1}
\end{equation}
where $\hat e$ is the identity element and $N$ is the order of the
cyclic group~$T_N$. The operator $\hat a$ induces the cyclic
permutation of all particles of the chain and, therefore, it acts on
the ``configuration vector'' $\vec{X}(t)$ as follows:
\[\hat a\vec{X}(t)\equiv
\hat a\{x_1(t),x_2(t),\dotsc,x_{N-1}(t),x_N(t)\}=
\{x_N(t),x_1(t),x_2(t),\dotsc,x_{N-1}(t)\}.\]

The full symmetry group of the monoatomic chain contains also the
inversion $\hat\i$, with respect to the center of the chain, which
acts on the vector $\vec{X}(t)$ in the following manner:
\[\hat\i\vec{X}(t)\equiv
\hat\i\{x_1(t),x_2(t),\dotsc,x_{N-1}(t),x_N(t)\}=
\{-x_N(t),-x_{N-1}(t),\dotsc,-x_2(t),-x_1(t)\}.\]

The complete set of all products $\hat a^k\hat\i$ of the pure
translations $\hat a^k$ ($k=0,1,2,\dotsc,N-1$) with the inversion
$\hat\i$ forms the so-called dihedral group $D_N$ which can be
written as the direct sum of two cosets $T_N$ and $T_N\cdot\hat\i$:
\begin{equation}
D_N=T_N\oplus T_N\cdot\hat\i. \label{eq4_2}
\end{equation}
The dihedral group is a non-Abelian group induced by two generators
($\hat a$ and $\hat\i$) with the following generating relations
\begin{equation}
\hat a^N=\hat e,\quad\hat\i^2=\hat e,\quad\hat\i\hat a=\hat
a^{-1}\hat\i.
 \label{eq4_3}
\end{equation}

We will consider different vibrational regimes in the FPU chains,
which can be determined by the \emph{specific forms} of the
configuration vector. Each of these regimes depends on $m$
independent parameters ($m\leq N$) and this number is the
\emph{dimension} of the given regime.

The simplest case of one-dimensional vibrational regimes represents
the so-called $\pi$-mode (zone boundary mode) \footnote{The dots in
$\vec{X}(t)$ denote that the displacement fragment which is given
explicitly must be repeated several times to form the full
displacement pattern corresponding to the given bush.}:
\begin{equation}
\vec{X}(t)=\{A(t),-A(t)~|~A(t),-A(t)~|~A(t),-A(t)~|~\dots\},
 \label{eq5}
\end{equation}
where $A(t)$ is a certain function of $t$. In our terminology, this
is the one-dimensional bush B$[\hat a^2,\hat\i]$ (see below about
notation of bushes of modes).

The vector
\begin{equation}
\vec{X}(t)=\{0,A(t),B(t),0,-B(t),-A(t)~|~\dots\},
 \label{eq6}
\end{equation}
represents a two-dimensional vibrational regime that is determined
by two time-dependent functions $A(t)$ and $B(t)$. This is the
two-dimensional bush B$[\hat a^6,\hat a\hat\i]$ (see \cite{FPU2}).

In general, for $m$-dimensional vibrational regime, we write
$\vec{X}(t)=\vec{C}(t)$, where the $N$-dimensional vector
$\vec{C}(t)$ depends on $m$ time-dependent functions only. Each
specific dynamical regime $\vec{C}(t)$, being an invariant manifold,
possesses its own symmetry group that is a subgroup of the parent
symmetry group $G_0=D_N$ of the chain in equilibrium.

\subsection{FPU-$\alpha$ chain with $N=4$ particles: existence of the bush
B$[\hat a^2,\hat\i]$}

Let us consider the above discussed equations for the simplest case
$N=4$. Dynamical equations (\ref{eq1}) read:
\begin{equation}
\begin{array}{l}
\ddot{x}_1=f(x_2-x_1)-f(x_1-x_4),\\
\ddot{x}_2=f(x_3-x_2)-f(x_2-x_1),\\
\ddot{x}_3=f(x_4-x_3)-f(x_3-x_2),\\
\ddot{x}_4=f(x_1-x_4)-f(x_4-x_3).
\end{array}
 \label{eq10}
\end{equation}

The symmetry group $G_0$ in the equilibrium state reads:
\[
G_0=D_4=[\hat a,\hat\i]=\{\hat e,\hat a,\hat a^2,\hat
a^3,\hat\i,\hat a\hat\i,\hat a^2\hat\i,\hat a^3\hat\i\}.
\]
Hereafter, we write generators of any symmetry group in square
brackets, while all its elements (if it is necessary) are given in
curly brackets.

The operators $\hat a$ and $\hat\i$ act on the configuration vector
$\vec{X}=\{x_1,x_2,x_3,x_4\}$ as follows
\[
\hat a\vec{X}=\{x_4,x_1,x_2,x_3\},\qquad
\hat\i\vec{X}=\{-x_4,-x_3,-x_2,-x_1\}.
\]
Therefore, we can associate the following matrices $\mathrm{M}(\hat
a)$ and $\mathrm{M}(\hat\i)$ of the \emph{mechanical representation}
with these generators:
\begin{equation}
\hat a\Rightarrow\mathrm{M}(\hat a)= \left(
\begin{array}{rrrr}
    0& 0& 0& 1\\
    1& 0& 0& 0\\
    0& 1& 0& 0\\
    0& 0& 1& 0\\
\end{array}\right),\qquad
\hat\i\Rightarrow\mathrm{M}(\hat\i)=\left(
\begin{array}{rrrr}
    0& 0& 0&-1\\
    0& 0&-1& 0\\
    0&-1& 0& 0\\
   -1& 0& 0& 0\\
\end{array}\right).
 \label{eq11}
\end{equation}
Their action on the configuration vector $\vec{X}$ is equivalent to
that of the operators $\hat a$ and $\hat\i$, respectively.

Let us now make the transformations of variables in the system
(\ref{eq10}) according to the action of the matrices (\ref{eq11}),
i.e.
\begin{equation}
\mathrm{M}(\hat a): x_1\rightarrow x_4,\quad x_2\rightarrow
x_1,\quad x_3\rightarrow x_2,\quad x_4\rightarrow x_3,
 \label{eq12}
\end{equation}
\begin{equation}
\mathrm{M}(\hat\i): x_1\rightarrow -x_4,\quad x_2\rightarrow
-x_3,\quad x_3\rightarrow -x_2,\quad x_4\rightarrow -x_1.
 \label{eq13}
\end{equation}
It is easy to check that both transformations (\ref{eq12}) and
(\ref{eq13}) produce systems of equations which are
\emph{equivalent} to the system (\ref{eq10}). Moreover, these
transformations act on the individual equations $u_j$ ($j=1,2,3,4$)
of the system (\ref{eq10}) exactly as on the components $x_j$
($j=1,2,3,4$) of the configuration vector $\vec{X}$. For example,
for the operator $\hat\i$ (or matrix $\mathrm{M}(\hat\i)$) we have
 \[
\hat\i u_1=-u_4,\quad\hat\i u_2=-u_3,\quad\hat\i
u_3=-u_2,\quad\hat\i u_4=-u_1.
 \]
It is obvious that a certain transposition of these equations
multiplied by $\pm1$ will indeed, produce a system fully identical
to the original system (\ref{eq10}). Thus, we are convinced that the
symmetry group $G_0=D_4$ of our chain in equilibrium turns out to be
the symmetry group (the group of invariance) of the dynamical
equations of this mechanical system.

Let us now consider the vibrational regime (\ref{eq5}), i.e.
$\pi$-mode, and check that it represents an invariant manifold for
the dynamical system (\ref{eq10}). Substituting
$x_1(t)=x_3(t)=A(t)$, $x_2(t)=x_4(t)=-A(t)$ into (\ref{eq10}), we
reduce these equations to one and the same equation of the form
\begin{equation}
\ddot{A}=f(-2A)-f(2A).
 \label{eq20}
\end{equation}
In the case of the FPU-$\alpha$ model this equation turns out to be
the equation of harmonic oscillator (for the FPU-$\beta$ model it
reduces to the Duffing equation). Indeed, for the FPU-$\alpha$
chain, we obtain from the Eq.(\ref{eq20}):
\begin{equation}
\ddot{A}+4A=0.
 \label{eq21}
\end{equation}
Using, for simplicity, the initial condition $A(0)=C_0$,
$\dot{A}(0)=0$, we get the following solution to Eq.~(\ref{eq21}):
\begin{equation}
A(t)=C_0\cos(2t).
 \label{eq22}
\end{equation}
Thus, the one-dimensional bush B$[\hat a^2,\hat\i]$ (or $\pi$-mode)
(\ref{eq5}) for the FPU-$\alpha$ chain, represents purely harmonic
dynamical regime
\begin{equation}
\vec{X}(t)=C_0\{\cos(2t),-\cos(2t)~|~\cos(2t),-\cos(2t)\}.
 \label{eq23}
\end{equation}

On the other hand, the invariant manifold
$\vec{X}(t)=\{A(t),-A(t),A(t),-A(t)\}$, corresponding to the bush
B$[\hat a^2,\hat\i]$, can be obtained with the aid of the
group-theoretical methods only, \emph{without} consideration of the
dynamical equations (\ref{eq10}). Let us discuss this point in more
detail.

At an arbitrary instant $t$, the displacement pattern
$\{A(t),-A(t),A(t),-A(t)\}$ possesses its own symmetry group
$G=D_2\subset G_0=D_4$. Indeed, this pattern is conserved under
inversion ($\hat\i$) and under shifting all particles by $2a$. The
latter procedure can be considered as a result of the action on the
chain by the operator $\hat a^2$. These two symmetry elements ($\hat
a^2$ and $\hat\i$) determine the dihedral group $D_2$ which is a
subgroup of order two of the original group $G_0=D_4$ \footnote{Let
us recall that the order $m$ of the subgroup $G$ in the group $G_0$
is determined by the equation
$m=\left\|G_0\right\|/\left\|G\right\|$, where $\left\|G_0\right\|$
and $\left\|G\right\|$ are numbers of elements in the groups $G_0$
and $G$, respectively.}.

It is obvious, that the old element $\hat a$ of the group $G_0$,
describing the chain in equilibrium, \emph{does not survive} in the
vibrational state described by the pattern (\ref{eq5}). Note that
this element ($\hat a$) transforms the regime (\ref{eq5}) into its
equivalent (but different!) form $\{-A(t),A(t),-A(t),A(t)\}$. In the
present paper, we will not discuss different equivalent forms of
bushes of modes (a detailed consideration of this problem can be
found in \cite{FPU2}). Thus, we encounter the \emph{reduction} of
symmetry $G_0=D_4\rightarrow G=D_2$ when we pass from the
equilibrium state to the vibrational state (\ref{eq5}) for the
considered mechanical system.

The dynamical regime (\ref{eq5}) represents the one-dimensional bush
consisting of only one mode ($\pi$-mode). We will denote it as
B$[G]$=B$[\hat a^2,\hat\i]:~\{A,-A,A,-A\}$. In square brackets, the
group of the bush symmetry is indicated by listing its generators
($\hat a^2$ and $\hat\i$, in our case), while the characteristic
fragment of the bush displacement pattern is presented next to the
colon. The bush symmetry group $G$ fully determines the form
(displacement pattern) of the bush B$[G]$ (see, for example,
\cite{PhysD,FPU2}). Indeed, in the case of the bush B$[\hat
a^2,\hat\i]$, it is easy to show that this form,
$\vec{X}=\{A(t),-A(t),A(t),-A(t)\}$, can be obtained as the general
solution to the following linear algebraic equation representing the
invariance of the configuration vector $\vec{X}$: $\hat
g_1\vec{X}=\vec{X}$, $\hat g_2\vec{X}=\vec{X}$, where $\hat g_1=\hat
a^2$ and $\hat g_2=\hat\i$ are the generators of the group $G$. In
our previous papers we often write these invariance conditions for
the bush B$[G]$ in the form
\begin{equation}
\hat G\vec{X}=\vec{X}.
 \label{eq30}
\end{equation}
It is very essential, that the invariant vector $\vec{X}(t)$, which
was found in such \emph{geometrical} (group-theoretical) manner,
turns out to be an invariant manifold for the considered dynamical
system \cite{PhysD}. Thus, we can obtain the symmetry-determined
invariant manifolds (bushes of modes) without any information on
interparticle interactions in the mechanical system.

\subsection{FPU-$\alpha$ chain with $N=4$ particles: stability of the bush
B$[\hat a^2,\hat\i]$\label{Sec2.3}}

We now turn to the question of the stability of the bush B$[\hat
a^2,\hat\i]$, representing a \emph{periodic} vibrational regime
$\vec{X}=\{A(t),-A(t),A(t),-A(t)\}$, with $A(t)=C_0\cos(2t)$.
According to the conventional prescription, we must linearize the
dynamical system (\ref{eq10}) in the \emph{infinitesimal vicinity}
of the given bush and then study the obtained system. For this goal,
let us write
\begin{equation}
\vec{X}(t)=\vec{C}(t)+\vec{\delta}(t),
 \label{eq40}
\end{equation}
where $\vec{C}=\{A(t),-A(t),A(t),-A(t)\}$ represents our bush, while
$\vec{\delta}(t)=\{\delta_1(t),\delta_2(t),\delta_3(t),\delta_4(t)\}$
is an infinitesimal vector. Substituting (\ref{eq40}) into
Eqs.\,(\ref{eq10}) and neglecting all terms nonlinear in
$\delta_j(t)$, we obtain the following linearized equations for the
FPU-$\alpha$ model:
\begin{equation}
\begin{array}{l}
\ddot{\delta}_1=[\delta_2-2\delta_1+\delta_4]-4A(t)\cdot[\delta_2-\delta_4],\\
\ddot{\delta}_2=[\delta_3-2\delta_2+\delta_1]+4A(t)\cdot[\delta_3-\delta_1],\\
\ddot{\delta}_3=[\delta_4-2\delta_3+\delta_2]-4A(t)\cdot[\delta_4-\delta_2],\\
\ddot{\delta}_4=[\delta_1-2\delta_4+\delta_3]+4A(t)\cdot[\delta_1-\delta_3].
 \label{eq41}
\end{array}
\end{equation}
The last system of equations can be written in the form
\begin{equation}
\ddot{\vec{\delta}}=\mathrm{J}(t)\cdot\vec{\delta},
 \label{eq42}
\end{equation}
where $\mathrm{J}(t)$ is the \emph{Jacobi} matrix for the system
(\ref{eq10}) calculated by the substitution of the vector
$\vec{X}=\{A(t),-A(t),A(t),-A(t)\}$. This matrix can be presented as
follows:
\begin{equation}
\mathrm{J}(t)=\mathrm{L}+4A(t)\cdot\mathrm{M},
 \label{eq43}
\end{equation}
where
\begin{equation}
\mathrm{L}=\left(
\begin{array}{rrrr}
  -2& 1& 0& 1\\
   1&-2& 1& 0\\
   0& 1&-2& 1\\
   1& 0& 1&-2
\end{array}\right)
,\qquad \mathrm{M}=\left(
\begin{array}{rrrrrr}
   0&-1& 0& 1\\
  -1& 0& 1& 0\\
   0& 1& 0&-1\\
   1& 0&-1& 0
\end{array}\right)
 \label{eq44}
\end{equation}
are two time-independent symmetric matrices.

It easy to check that matrices $\mathrm{L}$ and $\mathrm{M}$ commute
with each other:
$\mathrm{L}\cdot\mathrm{M}=\mathrm{M}\cdot\mathrm{L}$. Therefore,
there exists a time-independent orthogonal matrix $\mathrm{S}$ that
transforms the both matrices $\mathrm{L}$ and $\mathrm{M}$ to the
diagonal form:
$\mathrm{\tilde{S}}\cdot\mathrm{L}\cdot\mathrm{S}=\mathrm{L}_{dia}$,
$\mathrm{\tilde{S}}\cdot\mathrm{M}\cdot\mathrm{S}=\mathrm{M}_{dia}$
(here $\mathrm{\tilde{S}}$ is the transposed matrix with respect to
$\mathrm{S}$). In turn, it means that the Jacobi matrix
$\mathrm{J}(t)$ can be diagonalized at \emph{any time} $t$ by one
and the same time-independent matrix $\mathrm{S}$. Therefore, our
linearized system (\ref{eq42}) for the considered bush B$[\hat
a^2,\hat\i]$ can be decomposed into four \emph{independent}
differential equations.

Let us discuss how the above matrix $\mathrm{S}$ can be obtain with
the aid of the theory of irreducible representations of the symmetry
group $G$ (in our case $G=D_2$).

In Sec.\,\ref{Sec4}, we will consider a general method for obtaining
the matrix $\mathrm{S}$ which reduces the Jacobi matrix
$\mathrm{J}(t)$ to a block-diagonal form. This method uses the basis
vectors of irreducible representations of the group $G$, constructed
in the mechanical space of the considered dynamical system. In our
simplest case of the monoatomic chain with $N=4$ particles, this
method leads to the following result
\begin{equation}
\mathrm{S}=\frac{1}{2}
\left(
\begin{array}{rrrr}
  1& 1& 1& 1\\
  1& 1&-1&-1\\
  1&-1& 1&-1\\
  1&-1&-1& 1
\end{array}\right).
 \label{eq50}
\end{equation}
The rows of the matrix $\mathrm{S}$ from (\ref{eq50}) are simply the
characters of four one-dimensional irreducible representations
(irreps) –-- $\Gamma_1$, $\Gamma_2$, $\Gamma_3$, $\Gamma_4$ –-- of
the Abelian group $D_2$, because each of these irreps is contained
once in the decomposition of the mechanical representation of the
group $G=D_2$. Introducing new variables
$\vec{y}=\{y_1,y_2,y_3,y_4\}$ instead of the old variables
$\vec{\delta}=\{\delta_1(t),\delta_2(t),\delta_3(t),\delta_4(t)\}$
by the equation $\vec{y}=\mathrm{S}\cdot\vec{\delta}$ with
$\mathrm{S}$ from (\ref{eq50}), we arrive at the full splitting of
the linearized equations (\ref{eq41}) for the FPU-$\alpha$ model:
\begin{subequations}
\label{eq51}
\begin{eqnarray}
&&\ddot y_1=0,\label{eq51a}\\
&&\ddot y_2=-2[1+4A(t)]y_2,\label{eq51b}\\
&&\ddot y_3=-4y_3,\label{eq51c}\\
&&\ddot y_4=-2[1-4A(t)]y_4,\label{eq51d}
\end{eqnarray}
\end{subequations}
 where $A(t)=C_0\cos(2t)$.

With the aid of Eqs.\,(\ref{eq51}), we can find the stability
threshold in $C_0$ for loss of stability of the one-dimensional bush
B$[\hat a^2,\hat\i]$. Indeed, according to Eqs.\,(\ref{eq51}), the
variables $y_j(t)$ ($j=1,2,3,4$) are independent from each other,
and we can consider them in turn. Eq.\,(\ref{eq51a}) for $y_1(t)$
describes the uniform motion of the center of masses of our chain,
since it follows from the equations
$\vec{y}=\mathrm{S}\cdot\vec{\delta}$ that
$y_1(t)\sim(\delta_1(t)+\delta_2(t)+\delta_3(t)+\delta_4(t))$.
Therefore, considering \emph{vibrational} regimes only, we may
assume $y_1(t)\equiv0$.

If only $y_3(t)$ appears in the solution to the system (\ref{eq51}),
i.e. if $y_1(t)=0$, $y_2(t)=0$, $y_4(t)=0$, then we have from the
equation $\vec{\delta}=\tilde{\mathrm{S}}\cdot\vec{y}$ (note that
$\mathrm{S}$ is the orthogonal matrix and, therefore,
$\mathrm{S}^{-1}=\tilde{\mathrm{S}}$):
$\vec{\delta}(t)=\{y_3(t),-y_3(t),y_3(t),-y_3(t)\}$, where
$y_3(t)\sim\cos(2t)$. This solution leads only to deviations
``along'' the bush
$\vec{X}(t)=C_0\{\cos(2t),-\cos(2t),\cos(2t),-\cos(2t)\}$ and does
not signify instability.

Since $A(t)=C_0\cos(2t)$, Eq.~(\ref{eq51b}) reads
$\ddot{y}_2+[2+8C_0\cos(2t)]y_2=0$ and can be transformed to the
standard form of the Mathieu equation, as well as Eq.~(\ref{eq51d}).
Therefore, the stability threshold of the considered bush B$[\hat
a^2,\hat\i]$ for $N=4$ can be determined directly from the
well-known diagram of the regions of stable and unstable motion of
the Mathieu equation. In such a way we can find that critical value
$C_c$ for the amplitude $C_0$ of the given bush for which it loses
its stability is $C_c$ =$0.303$.

In conclusion, let us focus on the point that turns out to be very
important for proving the general theorem in Sec.\,\ref{Sec3}. The
system (\ref{eq41}) was obtained by linearizing the original system
(\ref{eq10}), near the dynamical regime
$\vec{C}(t)=\{A(t),-A(t),A(t),-A(t)\}$, and Eqs.~(\ref{eq10}) are
invariant with respect to the parent group $G_0=[\hat a,\hat\i]$.
Despite this fact, Eqs.~(\ref{eq41}) are invariant only with respect
to its \emph{subgroup} $G=\{\hat e,\hat a^2,\hat\i,\hat
a^2\hat\i\}\subset G_0=\{\hat e,\hat a,\hat a^2,\hat a^3,\hat\i,\hat
a\hat\i,\hat a^2\hat\i,\hat a^3\hat\i\}$: the element $\hat a\in
G_0$ (as well as $\hat a^3$, $\hat a\hat\i$, $\hat a^3\hat\i$) does
not survive as a result of the symmetry reduction $G_0\rightarrow
G$. Indeed, acting on Eqs.~(\ref{eq41}) by the operator $\hat g=\hat
a$, which transposes variables $\delta_j$ as follows
\begin{equation}
\delta_1\rightarrow\delta_4,\quad\delta_2\rightarrow\delta_1,\quad
\delta_3\rightarrow\delta_2,\quad\delta_4\rightarrow\delta_3,
 \label{eq70}
\end{equation}
we obtain the equations:
\begin{equation}
\begin{array}{l}
\ddot{\delta}_4=[\delta_1-2\delta_4+\delta_3]-4A(t)\cdot[\delta_1-\delta_3],\\
\ddot{\delta}_1=[\delta_2-2\delta_1+\delta_4]+4A(t)\cdot[\delta_2-\delta_4],\\
\ddot{\delta}_2=[\delta_3-2\delta_2+\delta_1]-4A(t)\cdot[\delta_3-\delta_1],\\
\ddot{\delta}_3=[\delta_4-2\delta_3+\delta_2]+4A(t)\cdot[\delta_4-\delta_2].
 \label{eq71}
\end{array}
\end{equation}
Obviously, this system is not equivalent to the system (\ref{eq41})!
(The equivalence between (\ref{eq41}) and (\ref{eq71}) can be
restored, if, besides cyclic permutation (\ref{eq70}) in
Eqs.~(\ref{eq41}), we add the artificial transformation
$A(t)\rightarrow -A(t)$).

What is the source of this phenomenon? The original nonlinear
dynamical system, which can be written as
$\ddot{\vec{X}}=\vec{F}(\vec{X})$, is invariant under the action of
the operator $\hat g=\hat a$. Being linearized, by the substitution
$\vec X(t)=\vec C(t)+\vec\delta(t)$ and neglecting all the nonlinear
in $\delta_j(t)$ terms, it becomes
\begin{equation}
\ddot{\vec\delta}= \left.\left(\frac{\partial \vec F}{\partial \vec
X}\right)\right|_{\vec{X}=\vec{C}}\cdot\vec\delta=\mathrm{J}\left[\vec
C(t)\right]\cdot\vec\delta,
 \label{eq72}
\end{equation}
where $\mathrm{J}\left[\vec C(t)\right]$ is the Jacobi matrix. The
latter system is also invariant under the action of the operator
$\hat g=\hat a$, but its transformation must be correctly written as
follows
\begin{equation}
\ddot{\vec\delta}=\hat g^{-1}\left.\left(\frac{\partial \vec
F}{\partial \vec X}\right)\right|_{\vec{X}=\hat g\vec{C}}\cdot\hat
g\vec\delta=\hat g^{-1}\mathrm{J}\left[\hat g\vec
C(t)\right]\cdot\hat g\vec\delta.
 \label{eq73}
\end{equation}
In other words, we have to replace the vector $\vec X$ in the Jacobi
matrix by a \emph{transformed} vector, $\hat g\vec C$, near which
the linearization is performed. Thus, we must write this matrix in
the form $\mathrm{J}\left[\hat g\vec C(t)\right]$ instead of
$\mathrm{J}\left[\vec C(t)\right]$. In our case, $\hat g\vec
C(t)\equiv\hat a\vec C(t)=\{-A(t),A(t),-A(t),A(t)\}$ and, therefore,
we indeed have to add the above mentioned artificial transformation
$A(t)\rightarrow -A(t)$).

On the other hand, dealing with the linearized system
$\ddot{\vec{\delta}}=\mathrm{J}(t)\cdot\vec{\delta}$, we
conventionally consider the Jacobi matrix $\mathrm{J}\left[\vec
C(t)\right]\equiv\mathrm{J}(t)$ as a fixed (but depending on $t$)
matrix which \emph{does not change} when the operator $\hat g=\hat
a$ acts on the system
$\ddot{\vec{\delta}}=\mathrm{J}(t)\cdot\vec{\delta}$
--- this operator acts on the vector $\vec\delta$ only! The fact is
that we try to split the system
$\ddot{\vec{\delta}}=\mathrm{J}(t)\cdot\vec{\delta}$ into some
subsystems using the \emph{traditional algebraic transformations} of
the old variables $\delta_j$. Indeed, we introduce new variables
$\vec\delta_{new}=\mathrm{S}\cdot\vec\delta$, where $\mathrm{S}$ is
a suitable time-independent orthogonal matrix, and then obtain the
new system
$\ddot{\vec{\delta}}_{new}=\left(\tilde{\mathrm{S}}\cdot\mathrm{J}(t)\cdot
\mathrm{S}\right)\cdot\vec{\delta}_{new}$ that decomposes into a
number of subsystems.

\subsection{Stability of the bush B$[\hat a^2,\hat\i]$ for the FPU-$\alpha$
chain with $N>4$ particles}

Linearizing the dynamical equations of the FPU-$\alpha$ chain with
$N=6$ in the vicinity of the bush B$[\hat a^2,\hat\i]$ ($\pi$-mode),
we obtain the following Jacobi matrix in Eq.~(\ref{eq42}):
\[\mathrm{J}(t)=\mathrm{L}+4A(t)\cdot \mathrm{M},\]
where
\[\mathrm{L}=\left(
\begin{array}{rrrrrr}
  -2& 1& 0& 0& 0& 1\\
   1&-2& 1& 0& 0& 0\\
   0&1 &-2& 1& 0& 0\\
   0& 0& 1&-2& 1& 0\\
   0& 0& 0& 1&-2& 1\\
   1& 0& 0& 0& 1&-2
\end{array}\right)
,\qquad \mathrm{M}=\left(
\begin{array}{rrrrrr}
   0&-1& 0& 0& 0& 1\\
  -1& 0& 1& 0& 0& 0\\
   0& 1& 0&-1& 0& 0\\
   0& 0&-1& 0& 1& 0\\
   0& 0& 0& 1& 0&-1\\
   1& 0& 0& 0&-1& 0
\end{array}\right)
.\] These two symmetric matrices, unlike the case $N=4$, \emph{do
not} commute with each other:
\[\mathrm{L}\cdot\mathrm{M}-\mathrm{M}\cdot
\mathrm{L}=8A(t)\left(
\begin{array}{rrrrrr}
   0& 0& 1& 0&-1& 0\\
   0& 0& 0&-1& 0& 1\\
  -1& 0& 0& 0& 1& 0\\
   0& 1& 0& 0& 0&-1\\
   1& 0&-1& 0& 0& 0\\
   0&-1& 0& 1& 0& 0
\end{array}\right)\neq0.\]
As a consequence, we cannot diagonalize both matrices $\mathrm{L}$
and $\mathrm{M}$ simultaneously, i.~e. with the aid of one and the
same orthogonal matrix $\mathrm{S}$. Therefore, it is impossible to
diagonalize the Jacobi matrix $\mathrm{J}(t)$ in equation
$\ddot{\vec{\delta}}=\mathrm{J}(t)\cdot\vec{\delta}$ for all time
$t$. In other words, there are no such matrix $\mathrm{S}$ that
completely splits the linearized system for the bush B$[\hat
a^2,\hat\i]$ for the chain with $N=6$ particles.

This difference between the cases $N=4$ and $N=6$ (generally, for
$N>4$) can be explained as follows. The group $G=[\hat a^2,\hat\i]$
of the considered bush, in fact, determines \emph{different} groups
for the cases $N=4$ and $N=6$. Indeed, for $N=4$ $\quad$ $[\hat
a^2,\hat\i] \equiv \{\hat E,\hat a^2,\hat\i,\hat a^2\hat\i\}=D_2$,
while for $N=6$ $\quad$ $[\hat a^2,\hat\i] \equiv \{\hat E,\hat
a^2,\hat a^4,\hat\i,\hat a^2\hat\i,\hat a^4\hat\i\}=D_3$. The latter
group ($D_3$) is non-Abelian ($\hat\i \hat a^4 = a^2\hat\i$), unlike
the group $D_2$ ($\hat\i \hat a^2 = a^2\hat\i$) and, as a
consequence, it possesses not only one-dimensional irreducible
representations, but two-dimensional irreps, as well. It will be
shown in Sec.\,\ref{Sec4}, that precisely this fact does not permit
us to split fully the above discussed linearized system
\footnote{Actually, this fact can be understood, if one takes into
account that the two-dimensional irrep contains two times in the
decomposition of the mechanical representation of the considered
chain.}.

In spite of this difficulty, we can simplify the linearized system
$\ddot{\vec{\delta}}=\mathrm{J}(t)\cdot\vec{\delta}$ considerably
with the aid of some group-theoretical methods, which are discussed
in the two following sections. Now, we only would like to present
the final result of the above splitting for the case $N=6$:
\begin{subequations}
\label{eq850}
\begin{eqnarray}
 &&\ddot{y}_1=-4y_1,\label{eq850a}\\
 &&\ddot{y}_2=0,\label{eq850b}\\ \nonumber\\
 &&\left\{
 \begin{array}{l}
  \ddot{y}_3+2y_3=P(t)y_5,\\
  \ddot{y}_5+2y_5=\bar{P}(t)y_3,
 \end{array}
 \right.\label{eq850c}\\ \nonumber\\
 &&\left\{
 \begin{array}{l}
  \ddot{y}_4+2y_4=P(t)y_6,\\
  \ddot{y}_6+2y_6=\bar{P}(t)y_4.
 \end{array}
 \right.\label{eq850d}
\end{eqnarray}
\end{subequations}
Here $P(t)=e^{\frac{i\pi}{3}}-4A(t)[1+e^{-\frac{i\pi}{3}}]$, while
$\bar{P}(t)$ is the complex conjugate function with respect to
$P(t)$. The two-dimensional subsystems (\ref{eq850c}) and
(\ref{eq850d}) can be reduces to the real form (\ref{eq5101}) (see,
Sec.\,\ref{Sec5.1}) by a certain linear transformation.

Note, that the stability of the $\pi$-mode (the bush B$[\hat
a^2,\hat\i]$) was discussed in a number of papers
[\onlinecite{Bud,Sand,Flach},\,
\onlinecite{FPU1,FPU2,PR,Yosh,Shin,CLL,AntiFPU}] by different
methods and with an emphasis on different aspects of this stability.
In particular, in our paper \cite{FPU1}, a remarkable fact was
revealed for the FPU-$\alpha$ chain: the stability threshold of the
$\pi$-mode is one and the same for interactions with \emph{all} the
other modes of the chain. (For other one-dimensional nonlinear
modes, for both the FPU-$\alpha$ and FPU-$\beta$ chains, the
stability thresholds, determined by interactions with different
modes, are essentially different \cite{FPU2}).

\section{The general theorem and its consequence}\label{Sec3}

We consider an $N$-degrees-of-freedom mechanical system that
described by $N$ autonomous differential equations
\begin{equation}
\ddot{\vec{X}}=\vec{F}(\vec{X}),
 \label{eq100}
\end{equation}
where the configuration vector
$\vec{X}=\{x_1(t),x_2(t),\dots,x_N(t)\}$ determines the deviation
from the equilibrium state $\vec{X}=\{0,0,\dots,0\}$, while
vector-function
$\vec{F}(\vec{X})=\{f_1(\vec{X}),f_2(\vec{X}),\dots,f_N(\vec{X})\}$
determines the right-hand-sides of the dynamical equations.

We assume that Eq.\,(\ref{eq100}) is invariant under the action of a
discrete symmetry group $G_0$ which we call ``the parent symmetry
group'' of our mechanical system. This means that for all $g\in G_0$
Eq.~(\ref{eq100}) is invariant under the transformation of variables
\begin{equation}
\tilde{\vec{X}}=\hat{g}\vec{X},
 \label{eq101}
\end{equation}
where $\hat{g}$ is the operator associated with the symmetry element
$g$ of the group $G_0$ by the conventional definition
\[\hat{g}\vec{X}=\{g^{-1}x_1(t),\dots,g^{-1}x_N(t)\}.\]

Using (\ref{eq100}) and (\ref{eq101}), one can write $\vec{X}=\hat
g^{-1}\tilde{\vec{X}}$, $\hat
g^{-1}\ddot{\tilde{\vec{X}}}=\vec{F}(\hat g^{-1}\tilde{\vec{X}})$,
and finally
\begin{equation}
\ddot{\tilde{\vec{X}}}=\hat g\vec{F}(\hat g^{-1}\tilde{\vec{X}}).
 \label{eq102}
\end{equation}
On the other hand, renaming $\vec{X}$ from Eq.~(\ref{eq100}) as
$\tilde{\vec{X}}$, one can write
$\ddot{\tilde{\vec{X}}}=\vec{F}(\tilde{\vec{X}})$. Comparing this
equation with Eq.~(\ref{eq102}), we obtain
$\vec{F}(\tilde{\vec{X}})=\hat g\vec{F}(\hat
g^{-1}\tilde{\vec{X}})$, or
\begin{equation}
\vec{F}(\hat g\vec{X})=\hat g\vec{F}(\vec{X}).
 \label{eq1500}
\end{equation}
This is the condition of invariance of the dynamical equations
(\ref{eq100}) under the action of the operator $\hat g$. It must
hold for all $g\in G_0$ (obviously, it is sufficient to consider
such equivalence only for the \emph{generators} of the group $G_0$).

Let $\vec{X}(t)=\vec{C}(t)$ be an $m$-dimensional specific dynamical
regime in the considered mechanical system that corresponds to the
bush B$[G]$ ($G\subseteq G_0$). This means that there exist some
functional relations between the individual displacements $x_i(t)$
($i=1,2,\dots,N$), and, as a result, the system (\ref{eq100})
reduces to $m$ ordinary differential equations in terms of the
independent functions (we denoted them by $A(t)$, $B(t)$, $C(t)$,
etc. in the previous section, see, for example,
Eqs.~(\ref{eq5},\ref{eq6})).

The vector $\vec{C}(t)$ is a general solution to the equation (see,
Eq.\,(\ref{eq30})) \[\hat{G}\vec{X}=\vec{X},\] where $G$ is the
symmetry group of the given bush B$[G]$ ($G\subseteq G_0$).

Now, we want to study the \emph{stability} of the dynamical regime
$\vec{C}(t)$, corresponding to the bush B$[G]$. To this end, we must
linearize the dynamical equations (\ref{eq100}) in a vicinity of the
given bush, or more precisely, in a vicinity of the vector
$\vec{C}(t)$. Let
\begin{equation}
\vec{X}=\vec{C}(t)+\vec{\delta}(t),
 \label{eq104}
\end{equation}
where $\vec{\delta}(t)=\{\delta_1(t),\dots,\delta_N(t)\}$ is an
infinitesimal $N$-dimensional vector. Substituting $\vec{X}(t)$ from
(\ref{eq104}) into (\ref{eq100}) and linearizing these equations
with respect to $\vec{\delta}(t)$, we obtain
\begin{equation}
\ddot{\vec{\delta}}=\mathrm{J}[\vec{C}(t)]\cdot\vec{\delta},
 \label{eq105}
\end{equation}
where $\mathrm{J}[\vec{C}(t)]$ is the Jacobi matrix of the system
(\ref{eq100}):
\[
\mathrm{J}[\vec{C}(t)]= \left\|\left.\frac{\partial f_i}{\partial
x_j}\right|_{\vec{X}=\vec{C}(t)}\right\|
 .\]

Now, we intend to prove the following
\begin{_theorem}
The matrix $\mathrm{J}[\vec{C}(t)]$ of the linearized dynamical
equations near a given bush B$[G]$, determined by the configuration
vector $\vec{C}(t)$, commutes with all matrices $\mathrm{M}(g)$
($g\in G$) of the mechanical representation of the symmetry group
$G$ of the considered bush:
\[\mathrm{M}(g)\cdot\mathrm{J}[\vec{C}(t)]=\mathrm{J}[\vec{C}(t)]\cdot\mathrm{M}(g).\]
 \label{theor1}
\end{_theorem}
\begin{proof}
As was already discussed in Sec.\,\ref{Sec2}, the original nonlinear
system $\ddot{\vec{X}}=\vec{F}(\vec{X})$ transforms into the system
$\ddot{\vec{X}}=\hat g^{-1}\vec{F}(\hat g\vec{X})$ under the action
of the operator $\hat g$ associated with the symmetry element $g\in
G_0$ of the parent group $G_0$. According to Eq.~(\ref{eq1500}), the
invariance of our system with respect to the operator $\hat g$ can
be written as follows:
\begin{equation}
\hat g^{-1}\vec{F}(\hat g\vec{X})=\vec{F}(\vec{X}).
 \label{eq150}
\end{equation}
On the other hand, the system $\ddot{\vec{X}}=\vec{F}(\vec{X})$,
linearized in the vicinity of the vector $\vec{X}=\vec{C}(t)$ reads
$\ddot{\vec{\delta}}=\mathrm{J}\left[\vec{C}(t)\right]\cdot\vec{\delta}$
(see Eq.~(\ref{eq105})). Under the action of the operator $\hat g$,
it transforms, according to Eq.~(\ref{eq73}), into the system
\begin{equation}
\ddot{\vec{\delta}}=\hat g^{-1}\mathrm{J}\left[\hat
g\vec{C}(t)\right]\cdot\hat g\vec{\delta}.
 \label{eq151}
\end{equation}

Let us now consider the mechanical representation $\Gamma$ of the
parent symmetry group $G_0$. To this end, we chose the ``natural''
basis $\vec{\Phi}=\{\vec{e}_1,\vec{e}_2,...,\vec{e}_N\}$ in the
space of all possible displacements of individual particles
(configuration space):
\begin{equation}
\vec{e}_1=\left(
\begin{array}{c}
   1\\
   0\\
   0\\
   \vdots\\
   0
\end{array}\right),\quad
\vec{e}_2=\left(
\begin{array}{c}
   0\\
   1\\
   0\\
   \vdots\\
   0
\end{array}\right),~\dots~,~
\vec{e}_N=\left(
\begin{array}{c}
   0\\
   0\\
   0\\
   \vdots\\
   1
\end{array}\right).
 \label{eq600}
\end{equation}
Acting by an operator $\hat g$ ($g\in G$) on the vector $\vec{e}_j$,
we can write
\begin{equation}
\hat g\vec{e}_j=\sum_{i=1}^{N}\mathrm M_{ij}(g)\cdot\vec{e}_i,\qquad
j=1,2,\dots,N.
 \label{eq6001}
\end{equation}
This equation associates the matrix $\mathrm
M(g)\equiv\left\|\mathrm M_{ij}\right\|$ with the operator $\hat g$
and, therefore, with the symmetry element $g\in G$:
\begin{equation}
g\Rightarrow\hat g\Rightarrow\mathrm M(g).
 \label{eq6002}
\end{equation}
The set of matrices $\mathrm M(g)$ corresponding to all $g\in G$
forms the mechanical representation $\Gamma$ for our system
\footnote{According to the traditional definition of the
$n$-dimensional matrix representation of the group $G$, a matrix
$\mathrm M(g)$ is associated with the element $g\in G$, if $\hat
g\vec{\Phi}=\tilde{\mathrm M}(g)\vec{\Phi}$. Here
$\vec{\Phi}=\{\vec{\phi}_1(\vec{r}),\vec{\phi}_2(\vec{r}),\dots,\vec{\phi}_N(\vec{r})\}$
is the set of basis vectors, $\hat g$ is the operator acting on the
vectors as $\hat
g\vec{\phi}_i(\vec{r})=\vec{\phi}_i(g^{-1}\vec{r})$, and
$\tilde{\mathrm M}(g)$ is the matrix transposed with respect to the
matrix $\mathrm M(g)$.}. As a consequence of this definition, the
equation
\begin{equation}
\hat g\vec{C}=\mathrm M(g)\cdot\vec{C}
 \label{eq6003}
\end{equation}
is valid for any vector $\vec{C}$ determined in the basis
(\ref{eq600}) as $\vec{C}=\sum_{k=1}^{N}C_k\vec{e}_k$.

Using Eq.~(\ref{eq6003}), we can rewrite the equation (\ref{eq151})
in terms of matrices $\mathrm{M}(g)\equiv\mathrm{M}_g$ ($g\in G_0$)
of the mechanical representation of the group $G_0$:
\begin{equation}
\ddot{\vec{\delta}}=\mathrm{M}^{-1}_g\cdot\mathrm{J}\left[\mathrm{M}_g
\vec{C}(t)\right]\cdot\mathrm{M}_g\cdot\vec{\delta}.
 \label{eq152}
\end{equation}
Therefore, the invariance of the system
$\ddot{\vec{\delta}}=\mathrm{J}\left[\vec{C}(t)\right]\cdot\vec{\delta}$
with respect of the operator $\hat g$ (matrix $\mathrm{M}_g$) can be
written as the following relation
\begin{equation}
\mathrm{M}^{-1}_g\cdot\mathrm{J}\left[\mathrm{M}_g\vec{C}(t)\right]
\cdot\mathrm{M}_g=\mathrm{J}\left[\vec{C}(t)\right].
 \label{eq153}
\end{equation}

Now, let us suppose that $g$ is an element of the symmetry group $G$
of a given bush B$[G]$ ($G\subseteq G_0$). By the definition, all
the elements of this group ($g\in G$) leave invariant the vector
$\vec{C}(t)$ that determines the displacement pattern of this bush
\begin{equation}
\hat g\vec{C}(t)=\mathrm{M}_g\cdot\vec{C}(t)=\vec{C}(t), \qquad g\in
G.
 \label{eq154}
\end{equation}
Taking into account this equation, we obtain from (\ref{eq153}) the
relation
\begin{equation}
\mathrm{M}^{-1}_g\cdot\mathrm{J}\left[\vec{C}(t)\right]\cdot\mathrm{M}_g
=\mathrm{J}\left[\vec{C}(t)\right],
 \label{eq155}
\end{equation}
which holds for each element $g$ of the symmetry group $G$ of the
considered bush.

Rewriting (\ref{eq155}) in the form
\begin{equation}
 \mathrm{J}\left[\vec{C}(t)\right]\cdot\mathrm{M}_g=
\mathrm{M}_g\cdot\mathrm{J}\left[\vec{C}(t)\right],
 \label{eq501}
\end{equation}
 we arrive at the conclusion of our Theorem: all the matrices
$\mathrm{M}_g$ of the mechanical representation of the group $G$
\emph{commute} with the Jacobi matrix
$\mathrm{J}\left[\vec{C}(t)\right]$ of the linearized (near the
given bush) dynamical equations
$\ddot{\vec{\delta}}=\mathrm{J}\left[\vec{C}(t)\right]\cdot\vec{\delta}$.
\end{proof}

In what follows, we will introduce a simpler notation for the Jacobi
matrix:
\begin{equation}
\mathrm{J}\left[\vec{C}(t)\right]\equiv\mathrm{J}(t).
 \label{eq405}
\end{equation}

\noindent\emph{Remark\/}. We have proved that all the matrices
$\mathrm{M}_g$ with $g\in G$ commute with the Jacobi matrix
$\mathrm{J}(t)$ of the system (\ref{eq105}). But if we take a
symmetry element $g\in G_0$ that is not contained in $G$ ($g\in
G_0\setminus G$), the matrix $\mathrm{M}_g$ corresponding to $g$
\emph{may not} commute with $\mathrm{J}(t)$. An example of such
noncommutativity and the source of this phenomenon were presented in
Sec.\,\ref{Sec2.3}.

\subsubsection*{Consequence of Theorem \ref{theor1}}

Taking into account Theorem \ref{theor1}, we can apply the
well-known \emph{Wigner} theorem \cite{Dob} to split the linearized
system $\ddot{\vec{\delta}}=\mathrm{J}(t)\cdot\vec{\delta}$ into a
certain number of independent subsystems. Indeed, according to this
theorem, the matrix ($\mathrm{J}(t)$, in our case) commuting with
all the matrices of a representation $\Gamma$ of the group $G$
(mechanical representation, in our case), can be reduced to a very
specific block-diagonal form. The dimension of each block of this
form is equal to $n_j\ast m_j$, where $n_j$ is the dimension of a
certain \emph{irreducible representation} (irrep) $\Gamma_j$ of the
group $G$ containing $m_j$ times in the reducible representation
$\Gamma$. Moreover, these blocks possess a particular structure
which will be considered in Sec.~\ref{Sec4}.

To implement this splitting explicitly one must pass from the old
basis $\vec{\Phi}_{old}=\{\vec{e}_1,\vec{e}_2,\dots,\vec{e}_N,\}$ of
the mechanical space to the new basis
$\vec{\Phi}_{new}=\{\vec{\phi}_1,\vec{\phi}_2,\dots,\vec{\phi}_N,\}$
formed by the complete set of the basis vectors $\vec{\phi}_k$
($k=1,2,\dots,N$) of all the irreps of the group $G$. If
$\vec{\Phi}_{new}=\mathrm{S}\cdot\vec{\Phi}_{old}$ then the unitary
transformation \footnote{Here $\mathrm{S}^+$ is the Hermite
conjugated matrix with respect to the matrix $\mathrm{S}$}
\begin{equation}
\mathrm{J}_{new}(t)=\mathrm{S}^+\cdot\mathrm{J}_{old}(t)\cdot\mathrm{S}
 \label{eq601}
\end{equation}
produces the above discussed block-diagonal matrix
$\mathrm{J}_{new}(t)$ of the linearized system
$\ddot{\vec{\delta}}_{new}=\mathrm{J}_{new}(t)\cdot\vec{\delta}_{new}$
(here $\vec{\delta}_{old}=\mathrm{S}\cdot\vec{\delta}_{new}$).

In the next section, we will search the basis vectors
$\vec{\phi}_i[\Gamma_j]$ ($i=1,2,\dots,n_j$) of each irreducible
representation $\Gamma_j$ in the form
\begin{equation}
\vec{\phi}_i[\Gamma_j]=\{x_{ij}^{(1)},x_{ij}^{(2)},\dots,x_{ij}^{(N)}\},
 \label{eq602}
\end{equation}
where $x_{ij}^{(k)},$ ($k=1,2,\dots,N$) determines the displacement
of the $k$-th particle corresponding to the $i$-th basis vector
$\vec{\phi}_i[\Gamma_j]$ of the $j$-th irrep $\Gamma_j$. Actually,
this means that we search $\vec{\phi}_i[\Gamma_j]$ as a
superposition of the old basis vectors $\vec{e}_k$ ($k=1,2,\dots,N$)
of the mechanical space (see (\ref{eq600})):
\begin{equation}
\vec{\phi}_i[\Gamma_j]=\sum_{k=1}^{N}x_{ij}^{(k)}\cdot\vec{e}_k.
 \label{eq603}
\end{equation}
If we find all the basis vectors $\vec{\phi}_i[\Gamma_j]$ in such a
form, the coefficients $x_{ij}^{(k)}$ are obviously the elements of
the matrix $\mathrm{S}$ that determines the transformation
$\vec{\Phi}_{new}=\mathrm{S}\cdot\vec{\Phi}_{old}$ from the old
basis $\vec{\Phi}_{old}=\{\vec{e}_k~|~k=1,2,\dots,N\}$ to the new
basis
$\vec{\Phi}_{new}=\{\vec{\phi}_i[\Gamma_j]~|~i=1,2,\dots,n_j;~j=1,2,\dots\}$.
Here $j=1,2,\dots$ are indices of the irreducible representations
that contribute to the reducible mechanical representation $\Gamma$.

Thus, finding all the basis vectors $\vec{\phi}_i[\Gamma_j]$ of the
irreps $\Gamma_j$ in the form (\ref{eq602}) provides us directly
with the matrix $\mathrm{S}$ that diagonalizes the Jacobi matrix
$\mathrm{J}(t)$ of the linearized system
$\ddot{\vec{\delta}}=\mathrm{J}(t)\cdot\vec{\delta}$.

\section{Stability analysis of dynamical regimes in monoatomic
chains\label{Sec4}}

\subsection{Setting up the problem and Theorem~\ref{theor2}}

In general, the study of stability of periodic and, especially,
quasiperiodic dynamical regimes in the mechanical systems with many
degrees of freedom presents considerable difficulties. Indeed, for
this purpose, we must integrate large linearized (near the
considered regime) system of differential equations with
time-dependent coefficients. In the case of periodic regime, one can
use the Floquet method requiring integration over only one
time-period to construct the monodromy matrix. But for quasiperiodic
regime this method is inapplicable, and one often needs to solve
system of great number of differential equations for very large time
intervals to reveal instability (especially, near the stability
threshold).

In such a situation, a decomposition (splitting) of the full
linearized system into a number of independent subsystems of small
dimensions proves to be very useful. Moreover, this decomposition
can provide valuable information on generalized degrees of freedom
responsible for the loss of stability of the given dynamical regime
for the first time. Let us note that the number of such ``critical''
degrees of freedom can frequently be rather small.

We want to illustrate the above idea with the case of $N$-particle
monoatomic chains for $N\gg 1$. Let us introduce the following
notation. The bush B$[G]$ with the symmetry group $G$ containing the
translational subgroup $[\hat a^m]$ will be denoted by B$[\hat
a^m,\dots]$, where dots stand for other generators of the group $G$.
(Note, that any $m$-dimensional bush can exist only for the chain
with $N$ divisible by $m$).

\begin{_theorem}
Linear stability analysis of any bush B$[\hat a^m,\dots]$ in the
$N$-degrees-of-freedom monoatomic chain can be reduced to stability
analysis of isolated subsystems of the second order differential
equations with time-dependent coefficients whose dimensions do not
exceed the integer number $m$.
 \label{theor2}
\end{_theorem}

\noindent\emph{Corollary\/}. If the bush dimension is $d$, one can
pass on to the subsystems of \emph{autonomous} differential
equations with dimensions not exceeding $(m+d)$.

Before proving these propositions we must consider the procedure of
constructing the basis vectors of the irreducible representations of
the translational group $T$.

\subsection{Basis vectors of irreducible representations of
the translational groups\label{Sec4.2}}

The basis vectors of irreducible representations of different
symmetry groups are usually obtained by the method of projection
operators \cite{Dob}, but, in our case, it is easier to make use of
the ``direct'' method based on the definition of the group
representation \footnote{We already use this method in our previous
papers (see, for example, \cite{FPU1})}.

Let $\Gamma$ be an $n$-dimensional representation (reducible or
irreducible) of the group $G$, while $V[\Gamma]$ be the invariant
subspace corresponding to this representation that determined by the
set $\vec{\Phi}$ of $N$-dimensional basis vectors $\vec{\phi}_j$
($j=1,\dots,n$):
\begin{equation}
\vec{\Phi}=\{\vec{\phi}_1,\vec{\phi}_2,\dots,\vec{\phi}_n\}.
 \label{eq200}
\end{equation}

Acting on any basis vector $\vec{\phi}_j$ by an operator $\hat g$
($g\in G$) and bearing in mind the invariance of the subspace
$V[\Gamma]$, we can represent the vector $\hat g\vec{\phi}_j$ as a
superposition of all basis vectors from (\ref{eq200}). In other
words,
\begin{equation}
\hat g\vec{\Phi}\equiv\{\hat g\vec{\phi}_1,\hat
g\vec{\phi}_2,\dots,\hat
g\vec{\phi}_n\}=\tilde{\mathrm{M}}(g)\vec{\Phi},
 \label{eq201}
\end{equation}
where $\mathrm{M}(g)$ is the matrix corresponding, in the
representation $\Gamma$, to the element $g$ of the group $G$. (In
Eq.~(\ref{eq201}) we use tilde as the symbol of matrix
transposition). Eq.~(\ref{eq201}) associates with any $g\in G$ a
certain $n\times n$ matrix $\mathrm{M}(g)$ and encapsulates the
definition of matrix representation
\begin{equation}
\Gamma=\{\mathrm{M}(g_1),\mathrm{M}(g_2),\dots~\}.
 \label{eq202}
\end{equation}

The above mentioned ``direct'' method is based precisely on this
definition. Let us use it to obtain the basis vectors of the
irreducible representations for the translational group
$T\equiv[\hat a^m]$. We will construct these vectors in the
mechanical space of the $N$-particle monoatomic chain and,
therefore, each vector $\vec{\phi}_j$ can be written as follows:
\begin{equation}
\vec{\phi}_j=\{x_1,x_2,\dots,x_N\},
 \label{eq203}
\end{equation}
where $x_i$ is a displacement of $i$-th particle from its
equilibrium.

The group $T\equiv[\hat a^m]$ represents a \emph{translational
subgroup} corresponding to the bush B$[G]=$B$[\hat a^m,\dots]$. For
$N$-particle chain \footnote{Note, the relation $N\mod m=0$ must
hold!}, $T\equiv[\hat a^m]$ is a subgroup of the order $k=N/m$ of
the full translational group $T_N\equiv[\hat a]$, and we can write
the complete set of its elements as follows:
\begin{equation}
T_k=\{\hat e,\hat a^m,\hat a^{2m},\hat a^{3m},\dots,\hat
a^{(k-1)m}\}\qquad(\hat a^{km}=\hat a^N=\hat e).
 \label{eq210_}
\end{equation}

Being cyclic, the group $T_k$ from (\ref{eq210_}) possesses only
one-dimensional irreps, and their total number is equal to the order
($k=N/m$) of this group.

Below, for simplicity, we consider the case $m=3$ and $N=12$. The
generalization to the case of arbitrary values of $m$ and $N$ turns
out to be trivial.

As it is well-known, the one-dimensional irreps $\Gamma_i$ of the
$k$-order cyclic group can be constructed with the aid of $k$-degree
roots of $1$ and, therefore, for our case $N=12$, $m=3$, $k=4$, we
obtain the irreducible representations listed in Table~\ref{table1}.
\begin{table}
\caption{Irreducible representations of the cyclic group
$T_4$\label{table1}}
\begin{ruledtabular}
\begin{tabular}{c|cccc}
       &$\hat e$&$\hat g$&$\hat g^2$&$\hat g^3$\\
\hline
 $\Gamma_1$& $1$& $1$& $1$& $1$\\
 $\Gamma_2$& $1$& $i$&$-1$&$-i$\\
 $\Gamma_3$& $1$&$-1$& $1$&$-1$\\
 $\Gamma_4$& $1$&$-i$&$-1$& $i$
\end{tabular}
\end{ruledtabular}
\end{table}

In accordance with the definition (\ref{eq201}), the basis vector
$\vec{\phi}$ of the one-dimensional irrep $\Gamma$, for which
$\mathrm{M}(g)=\gamma$, must satisfy the equation
\begin{equation}
\hat g\vec{\phi}=\gamma \vec{\phi}.
 \label{eq210}
\end{equation}
In our case, $\hat g=\hat a^3$, this equation can be written as
follows:
\begin{equation}
\begin{array}{l}
\hat
g\vec{\phi}\equiv\{x_{10},x_{11},x_{12}|x_1,x_2,x_3|x_4,x_5,x_6|x_7,x_8,x_9\}=\\
\gamma\{x_1,x_2,x_3|x_4,x_5,x_6|x_7,x_8,x_9|x_{10},x_{11},x_{12}\}.
 \label{eq211}
\end{array}
\end{equation}
Here $\gamma=1,i,-1,-i$ for the irreps $\Gamma_1$, $\Gamma_2$,
$\Gamma_3$ and $\Gamma_4$, respectively. Equating the sequential
components of both sides of Eq.~(\ref{eq211}), we obtain the
\emph{general solution} to the equation $\hat g\vec{\phi}=\gamma
\vec{\phi}$ that turns out to depend on three arbitrary constants,
say, $x$, $y$ and $z$:
\begin{equation}
\begin{array}{l}
\vec{\phi}=\{x,y,z|\gamma^{-1}x,\gamma^{-1}y,\gamma^{-1}z|\gamma^{-2}x,\gamma^{-2}y,
\gamma^{-2}z|\gamma^{-3}x,\gamma^{-3}y,\gamma^{-3}z\}=\\
x\{1,0,0|\gamma^{-1},0,0|\gamma^{-2},0,0|\gamma^{-3},0,0\}+\\
y\{0,1,0|0,\gamma^{-1},0|0,\gamma^{-2},0|0,\gamma^{-3},0\}+\\
z\{0,0,1|0,0,\gamma^{-1}|0,0,\gamma^{-2}|0,0,\gamma^{-3}\}.
 \label{eq212}
\end{array}
\end{equation}
Here we write the vector $\vec{\phi}$ as the superposition (with
coefficients $x$, $y$, $z$) of \emph{three basis vectors}. It means
that the irrep $\Gamma$ is contained \emph{thrice} in the
decomposition of the mechanical representation into irreducible
representations of the group $G=[\hat a ^3]$.

This result can be generalized to the case of arbitrary $N$ and $m$
in trivial manner: each irrep of the group $G=[\hat a ^m]$ enters
\emph{exactly} $m$ \emph{times} into the decomposition of the
mechanical representation for $N$-particle chain, and the rule for
constructing $m$ appropriate basis vectors is fully obvious from
Eq.~(\ref{eq212}).

\subsection{Proof of Theorem\,\ref{theor2}}

\begin{proof}
The basis vectors of all irreps $\Gamma_i$, listed for the case
$N=12$, $m=3$ in Table\,\ref{table1}, can be obtained from
(\ref{eq212}) setting $\gamma=1,i,-1,-i$, respectively (these values
are one-dimensional matrices corresponding in $\Gamma_i$
($i=1,2,3,4$) to the generator $\hat g\equiv\hat a^3$).

Let us write the above basis vectors sequentially, as it is done in
Table\,\ref{table2}, and prove that $12\times12$ matrix, determined
by this table, is precisely the matrix $\mathrm S$ that splits the
linearized dynamical equations
$\ddot{\vec{\delta}}=\mathrm{J}(t)\cdot\vec{\delta}$ for the
considered case. In Table\,\ref{table2}, we denote the basis vectors
$\vec{\phi}_j(\Gamma_i)$ by the symbol of the irrep $\Gamma_i$
($i=1,2,3,4$) and the number $j=1,2,3$ of the basis vector of this
irrep. The normalization factor $\left(\frac{1}{2}\right)$ must be
associated with each row of this table to produce the normalized
basis vectors (because of this fact, we mark the rows as
$2\cdot\vec{\phi}_j(\Gamma_i)$ in the last column of
Table~\ref{table2}).

\begin{table}
\caption{Basis vectors of the irreducible representations of the
cyclic group $T_4$\label{table2}}
\begin{ruledtabular}
\begin{tabular}{ccc|ccc|ccc|ccc||l}
 $\delta_1$&$\delta_2$&$\delta_3$&$\delta_4$&$\delta_5$&$\delta_6$&
 $\delta_7$&$\delta_8$&$\delta_9$&$\delta_{10}$&$\delta_{11}$&
 $\delta_{12}$&\\
 $x_1$&$x_2$&$x_3$&$x_4$&$x_5$&$x_6$&$x_7$&$x_8$&$x_9$&
 $x_{10}$&$x_{11}$&$x_{12}$&\\
\hline\hline
$1$& $0$& $0$& $1$& $0$& $0$& $1$& $0$& $0$& $1$& $0$& $0$&
$2\cdot\vec{\phi}_1(\Gamma_1)$\\
$0$& $1$& $0$& $0$& $1$& $0$& $0$& $1$& $0$& $0$& $1$& $0$&
$2\cdot\vec{\phi}_2(\Gamma_1)$\\
$0$& $0$& $1$& $0$& $0$& $1$& $0$& $0$& $1$& $0$& $0$& $1$&
$2\cdot\vec{\phi}_3(\Gamma_1)$\\
\hline
$1$& $0$& $0$&$-i$& $0$& $0$&$-1$& $0$& $0$& $i$& $0$& $0$&
$2\cdot\vec{\phi}_1(\Gamma_2)$\\
$0$& $1$& $0$& $0$&$-i$& $0$& $0$&$-1$& $0$& $0$& $i$& $0$&
$2\cdot\vec{\phi}_2(\Gamma_2)$\\
$0$& $0$& $1$& $0$& $0$&$-i$& $0$& $0$&$-1$& $0$& $0$& $i$&
$2\cdot\vec{\phi}_3(\Gamma_2)$\\
\hline
$1$& $0$& $0$&$-1$& $0$& $0$& $1$& $0$& $0$&$-1$& $0$& $0$&
$2\cdot\vec{\phi}_1(\Gamma_3)$\\
$0$& $1$& $0$& $0$&$-1$& $0$& $0$& $1$& $0$& $0$&$-1$& $0$&
$2\cdot\vec{\phi}_2(\Gamma_3)$\\
$0$& $0$& $1$& $0$& $0$&$-1$& $0$& $0$& $1$& $0$& $0$&$-1$&
$2\cdot\vec{\phi}_3(\Gamma_3)$\\
\hline
$1$& $0$& $0$& $i$& $0$& $0$&$-1$& $0$& $0$&$-i$& $0$& $0$&
$2\cdot\vec{\phi}_1(\Gamma_4)$\\
$0$& $1$& $0$& $0$& $i$& $0$& $0$&$-1$& $0$& $0$&$-i$& $0$&
$2\cdot\vec{\phi}_2(\Gamma_4)$\\
$0$& $0$& $1$& $0$& $0$& $i$& $0$& $0$&$-1$& $0$& $0$&$-i$&
$2\cdot\vec{\phi}_3(\Gamma_4)$
\end{tabular}
\end{ruledtabular}
\end{table}

Obviously, we can use the matrix $\mathrm S$ from
Table\,\ref{table2} (the rows of this matrix are the basis vectors
of all the irreps of the group $T_4$) not only for the action on the
vectors in the $\vec{X}$-space of the full nonlinear system, but on
the vectors in the $\vec{\delta}$-space of the linearized system
$\ddot{\vec{\delta}}=\mathrm{J}(t)\cdot\vec{\delta}$, as well. It is
essential that in the latter case the matrix $\mathrm S$ reduces the
Jacobi matrix $\mathrm{J}(t)$ to a certain block-diagonal form.
Indeed, as was shown in Theorem\,\ref{theor1}, the matrix
$\mathrm{J}(t)$ commutes with all the matrices of the mechanical
representation of the bush symmetry group. Therefore, according to
Wigner theorem \cite{Dob}, it can be reduced, using unitary
transformation by the matrix $\mathrm S$, to the block-diagonal form
with blocks whose dimension is equal to $n_j\cdot m_j$. Here $n_j$
is the dimension of the irrep $\Gamma_j$, while $m_j$ is the number
of times that this irrep enters into the decomposition of the
mechanical representation (constructed, in our case, in the
$\vec{\delta}$-space).

In Sec.\,\ref{Sec4.2}, we have shown that for the translational
group $T=[\hat a^m]$ all $n_j=1$ and all $m_j=m$. Therefore, the
above matrix $\mathrm S$ decomposes the Jacobi matrix
$\mathrm{J}(t)$ into blocks whose dimension is equal to $m$. As a
consequence of this decomposition, the system
$\ddot{\vec{\delta}}=\mathrm{J}(t)\cdot\vec{\delta}$ splits into
$k=N/m$ independent subsystems, $L_j$ ($j=1,2,\dots,k$), each
consisting of $m$ differential equations of the second order. The
coefficients of these equations are time-dependent functions, and
this time dependence is determined by the functions $A(t)$, $B(t)$,
$C(t)$ etc., entering into the bush displacement pattern (see, for
example, (\ref{eq5},\ref{eq6})). For the one-dimensional bushes, the
coefficients of the above subsystems $L_j$ turn out to be periodic
functions with identical period, while for the many-dimensional
bushes they possess different periods (such bushes describe
quasiperiodic motion).

In general, it is impossible to obtain the explicit form of the
functions $A(t)$, $B(t)$, $C(t)$ etc., determining the bush
displacement pattern. Therefore, we can add the bush dynamical
equations to the $m$ differential equations of each subsystems
$L_j$. These additional $d$ equations determine the functions
$A(t)$, $B(t)$, $C(t)$ etc. \emph{implicitly}, where $d$ is the
dimension of the considered bush B$[\hat a^m,\dots]$.

On the other hand, we can give the following estimate for the bush
dimension $d$:
\begin{equation}
d\leq m.
 \label{eq260}
\end{equation}
Here $m$ is the index of the translational symmetry of the bush
B$[\hat a^m,\dots]$ (it determines the ratio between the size of the
primitive cell in the vibrational state and in the equilibrium).
Indeed, the bush displacement pattern can be found as the solution
to the equation $\hat{G}\vec{X}=\vec{X}$. If we take into account
only translational symmetry group of the bush B$[\hat a^m,\dots]$,
i.e. $G=[\hat a^m]$, this equation reduces to the equation $\hat
g\vec{\phi}=\vec{\phi}$ ($\hat g=\hat a^m$) for the basis vector
$\vec{\phi}$ of the \emph{identity} irrep ($\gamma=1$ in
(\ref{eq210})) of the group $T=[\hat a^m]$. As it has been already
shown in Sec.\,\ref{Sec4.2}, such vector $\vec{\phi}$ depends on
exactly $m$ arbitrary parameters. But some additional symmetry
elements, denoted by dots in the bush symbol B$[\hat a^m,\dots]$,
can lead to a decrease in the number $m$ of the above parameters.
For example, the bush B$[\hat a^3,\hat\i]$ turns out to be
one-dimensional, i.e., in this case, $d=1$, while $m=3$. Even the
\emph{vibrational} bush B$[\hat a^3]$ turns out to be
two-dimensional ($d=2$), if the condition of immobility of the mass
center is taken into account. Thus, for all cases, $d\leq m$, and we
can state that $m$ equations of each $L_j$, extended by $d$
additional equations of the given bush, provide us with $k=N/m$
independent subsystems $\tilde{L}_j$ of ($m+d$) \emph{autonomous}
differential equations.

Taking into account the additional bush symmetry elements, denoted
by dots in the symbol B$[\hat a^m,\dots]$, leads not only to
reducing the bush dimension, but to a further splitting of the above
discussed subsystems $L_j$ (we consider this point in the next
section). Thus, the linear stability analysis of the bush B$[\hat
a^m,\dots]$ in the $N$-particle chain indeed reduces to studying
stability of individual subsystems whose dimension does not exceed
$(m+d)$. This is the conclusion of Theorem~\ref{theor2} and, thus,
we have completed the proof.
\end{proof}

\subsection{Example 1: splitting the linearized system for the bush
 B$[\hat a^3]$\label{Sec4.4}}

We consider the splitting of the linearized system
$\ddot{\vec{\delta}}=\mathrm{J}(t)\cdot\vec{\delta}$ for the bush
B$[\hat a^3]$ in a chain with $N=12$ particles. The original
nonlinear system, for this case, reads
\begin{equation}
\begin{array}{l}
\ddot{x}_i=f(x_{i+1}-x_i)-f(x_i-x_{i-1}),\\
i=1,2,\dots,12\qquad(x_0=x_{12},\quad x_{13}=x_1).
 \label{eqk40}
\end{array}
\end{equation}
The displacement pattern of the bush B$[\hat a^3]$, obtained from
the equation $\hat a^3\vec{X}=\vec{X}$ reads:
\begin{equation}
\vec{X}=\{x(t),y(t),z(t)~|~x(t),y(t),z(t)~|~x(t),y(t),z(t)~|~x(t),y(t),z(t)\}.
 \label{eqk41}
\end{equation}
Substituting this form of vibrational pattern into (\ref{eqk40}), we
obtain three differential equations for the functions $x(t)$,
$y(t)$, $z(t)$ (all the other equations of (\ref{eqk40}) turn out to
be equivalent to these equations):
\begin{equation}
\begin{array}{l}
\ddot{x}=f(y-x)-f(x-z),\\
\ddot{y}=f(z-y)-f(y-x),\\
\ddot{z}=f(x-z)-f(z-y).
 \label{eqk42}
\end{array}
\end{equation}
The linearization of the Eqs.~(\ref{eqk40}) near the dynamical
regime determined by (\ref{eqk41}) leads to the system
\begin{equation}
\ddot{\vec{\delta}}=\mathrm{J}(t)\cdot\vec{\delta}
 \label{eqk43}
\end{equation}
with the following Jacobi matrix:
\begin{equation}
\mathrm{J}(t)=\left(
\begin{array}{cccccccccccc}
  \alpha& A&0&0&0&0&0&0&0&0&0&B\\
  A&\beta&C&0&0&0&0&0&0&0&0&0\\
  0& C&\gamma&B&0&0&0&0&0&0&0&0\\
  0&0&B&\alpha&A&0&0&0&0&0&0&0\\
  0&0&0&A&\beta&C&0&0&0&0&0&0\\
  0&0&0&0&C&\gamma&B&0&0&0&0&0\\
  0&0&0&0&0&B&\alpha&A&0&0&0&0\\
  0&0&0&0&0&0&A&\beta&C&0&0&0\\
  0&0&0&0&0&0&0&C&\gamma&B&0&0\\
  0&0&0&0&0&0&0&0&B&\alpha&A&0\\
  0&0&0&0&0&0&0&0&0&A&\beta&C\\
  B&0&0&0&0&0&0&0&0&0&C&\gamma\\
\end{array}\right),
 \label{eqk44}
\end{equation}
where
\begin{equation}
\begin{array}{l}
  A(t)=f'[y(t)-x(t)],\\
  B(t)=f'[x(t)-z(t)],\\
  C(t)=f'[z(t)-y(t)],\\
  \alpha(t)=-[A(t)+B(t)],\\
  \beta(t)=-[A(t)+C(t)],\\
  \gamma(t)=-[B(t)+C(t)].
\end{array}
 \label{eqk44a}
\end{equation}

Using Table\,\ref{table2}, the matrix $\mathrm{S}$ that splits up
the system (\ref{eqk43}) can be written as follows:
\begin{equation}
\mathrm{S}=\frac{1}{2}\left(
\begin{array}{rrrr}
  \mathrm{I}&  \mathrm{I}&  \mathrm{I}&  \mathrm{I}\\
  \mathrm{I}&-i\mathrm{I}& -\mathrm{I}& i\mathrm{I}\\
  \mathrm{I}& -\mathrm{I}&  \mathrm{I}& -\mathrm{I}\\
  \mathrm{I}& i\mathrm{I}& -\mathrm{I}&-i\mathrm{I}
\end{array}\right),
 \label{eqk45}
\end{equation}
where $\mathrm{I}$ is the $3\times3$ identity matrix
 \[
\mathrm{I}=\left(
\begin{array}{rrr}
  1& 0& 0\\
   0&1& 0\\
   0& 0&1
\end{array}\right).
 \]

With the aid of the unitary transformation
\begin{equation}
\mathrm{J}_{new}(t)=\mathrm{S}^+\cdot\mathrm{J}(t)\cdot\mathrm{S},
 \label{eqk46}
\end{equation}
we obtain
\begin{equation}
\mathrm{J}_{new}(t)=\left(
\begin{array}{cccc}
  \mathrm{D}_1& 0& 0& 0\\
  0&\mathrm{D}_2& 0& 0\\
  0& 0&\mathrm{D}_3& 0\\
  0& 0& 0&\mathrm{D}_4
\end{array}\right),
 \label{eqk47}
\end{equation}
where
\begin{equation}
\mathrm{D}_k=\left(
\begin{array}{ccc}
  -(A+B)& A& \gamma_kB \\
  A&-(A+C)&C\\
  \bar{\gamma}_kB& C&-(B+C)
\end{array}\right),\quad(k=1,2,3,4),
 \label{eqk48}
\end{equation}
with $\gamma_1=1$, $\gamma_2=i$, $\gamma_3=-1$, $\gamma_4=-i$
($\bar{\gamma}_k$ is the complex conjugate value of $\gamma_k$).

This means that the linear transformation
\begin{equation}
\vec{\delta}=\mathrm{S}\cdot\vec{\delta}_{new}
 \label{eqk49}
\end{equation}
reduces the old equations (\ref{eqk43}) to the following form
\begin{equation}
\ddot{\vec{\delta}}_{new}=\mathrm{J}_{new}(t)\cdot\vec{\delta}_{new}
 \label{eqk50}
\end{equation}
with block-diagonal matrix
$\mathrm{J}_{new}(t)=\mathrm{S}^+\cdot\mathrm{J}(t)\cdot\mathrm{S}$
determined by Eqs.~(\ref{eqk47},\ref{eqk48}).

Assuming
 \[
\vec{\delta}_{new}=\{\delta_1^{(1)},\delta_2^{(1)},\delta_3^{(1)}|
\delta_1^{(2)},\delta_2^{(2)},\delta_3^{(2)}|
\delta_1^{(3)},\delta_2^{(3)},\delta_3^{(3)}|
\delta_1^{(4)},\delta_2^{(4)},\delta_3^{(4)}\},
 \]
we can present (\ref{eqk50}) in a more explicit form:
\begin{equation}
\begin{array}{l}
  \ddot{\delta}_1^{(k)}=-(A+B)\delta_1^{(k)}+A\delta_2^{(k)}
  +\gamma_kB\delta_3^{(k)}, \\
  \ddot{\delta}_2^{(k)}=A\delta_1^{(k)}-(A+C)\delta_2^{(k)}
  +C\delta_3^{(k)}, \\
  \ddot{\delta}_3^{(k)}=\bar{\gamma}_kB\delta_1^{(k)}+C\delta_2^{(k)}
  -(B+C)\delta_3^{(k)},
\end{array}
 \label{eqk60}
\end{equation}
where $\gamma_k=1,i,-1,-i$ for $k=1,2,3,4$, respectively.

Thus, we obtain four independent $3\times3$ systems of linear
differential equations with time-dependent coefficients $A(t)$,
$B(t)$ and $C(t)$, which are determined by Eqs.~(\ref{eqk44a}).

Let us write down these equations for the FPU-$\alpha$ chain. For
this case, the function $f(x)$ in (\ref{eqk40}) reads $f(x)=x+x^2$.
Therefore, $f'(x)=1+2x$, and we obtain from (\ref{eqk44a})
 \[
 \begin{array}{l}
 A(t)=1+2[y(t)-x(t)],\\
 B(t)=1+2[x(t)-z(t)],\\
 C(t)=1+2[z(t)-y(t)].
\end{array}
 \]
Substituting these functions into (\ref{eqk60}) one can finally
obtain the following equations for the FPU-$\alpha$ chain:
 \[
 \ddot{\vec{\delta}}^{(k)}=\mathrm{J}_k(t)\cdot\vec{\delta}^{(k)},
 \]
where
\begin{equation}
\mathrm{J}_k(t)=\left(
\begin{array}{ccc}
  -2(1+y-z)& 1+2(y-x)& \gamma_k[1+2(x-z)] \\
  1+2(y-x)&-2(1+z-x)&1+2(z-y)\\
  \bar{\gamma}_k[1+2(x-z)]& [1+2(z-y)]&-2(1+x-y)
\end{array}\right),\quad(k=1,2,3,4).
 \label{eqk61}
\end{equation}
Here $\gamma_1=1$, $\gamma_2=i$, $\gamma_3=-1$, $\gamma_4=-i$ and
$x\equiv x(t)$, $y\equiv y(t)$, $z\equiv z(t)$ are functions
determined by the dynamical equations of the bush B$[\hat a^3]$ :
\begin{equation}
\begin{array}{l}
  \ddot{x}=(y-2x+z)\cdot(1+y-z), \\
  \ddot{y}=(z-2y+x)\cdot(1+z-x), \\
  \ddot{z}=(x-2z+y)\cdot(1+x-y).
\end{array}
 \label{eqk62}
\end{equation}
These equations can be obtained from (\ref{eqk42}) taking into
account the relation $f(x)=x+x^2$ for the case of the FPU-$\alpha$
model.

\noindent\emph{Remark\/}. According to (\ref{eqk41}), (\ref{eqk42})
(see, also (\ref{eqk62})), the vibrational bush B$[\hat a^3]$ is
three-dimensional. However, it actually turns out to be a
\emph{two-dimensional} bush. Indeed, there is no \emph{onsite}
potential in the FPU-$\alpha$ chains and, therefore, the
conservation law of the total momentum of such system holds.
Assuming that the center of masses is fixed, we obtain an additional
relation $x(t)+y(t)+z(t)=0$ which reduces the dimension of the bush
B$[\hat a^3]$ from $3$ to $2$.

\subsection{Further decomposition of linearized systems based on higher symmetry
groups\label{Sec4.5}}

Up to this point, we have discussed the decomposition of the
linearized system
$\ddot{\vec{\delta}}=\mathrm{J}(t)\cdot\vec{\delta}$ using only the
\emph{translational} part of the bush symmetry group. In general,
one can arrive at a more detailed splitting, if one takes into
account the additional bush symmetries.

\subsubsection{Example 2: Splitting of the linearized system for the bush
B$[\hat a^3,\hat\i]$\label{Sec4.5.1}}

Let us consider the decomposition of the linearized system for the
bush B$[\hat a^3,\hat\i]$ in the case of an arbitrary monoatomic
chain. Since translational part of the symmetry group $G=[\hat
a^3,\hat\i]$, which turns out to be the dihedral group, is the same
as that of the early considered bush B$[\hat a^3]$, we can take
advantage of all the results obtained in Sec.\,\ref{Sec4.4} and add
only some restrictions originating from the presence of the
additional generator $\hat\i$ of the group $[\hat a^3,\hat\i]$.

Substituting the vector $\vec{X}(t)$ in the form (\ref{eqk41}) into
the equation $\hat\i\vec{X}(t)=\vec{X}(t)$, we obtain
$z(t)\equiv-x(t)$, $y(t)\equiv-y(t)$ and, therefore, $y(t)\equiv0$.
The displacement pattern for the bush B$[\hat a^3,\hat\i]$ then can
be written as follows:
\begin{equation}
\vec{X}(t)=\{x(t),0,-x(t)~|~x(t),0,-x(t)~|~x(t),0,-x(t)~|~x(t),0,-x(t)\}.
 \label{eqk70}
\end{equation}
Thus, the bush B$[\hat a^3,\hat\i]$ turns out to be
\emph{one-dimensional}.

As a result of the substitution $z(t)\equiv-x(t)$, $y(t)\equiv0$,
three equations (\ref{eqk42}) reduce to only one equation
\begin{equation}
\ddot{x}=f(-x)-f(2x).
 \label{eqk71}
\end{equation}
For the FPU-$\alpha$ chain (see Eqs.~(\ref{eqk62})), this equation
transforms to
\begin{equation}
\ddot{x}+3x+3x^2=0.
 \label{eqk72}
\end{equation}

Unlike the purely translational group $[\hat a^3]$, of the
three-dimensional bush B$[\hat a^3]$, the symmetry group $[\hat
a^3,\hat\i]$ of the one-dimensional bush B$[\hat a^3,\hat\i]$ is the
dihedral group with another sets of the irreps and basis vectors. It
can be shown that taking into account that $[\hat a^3,\hat\i]$ is
the supergroup with respect to group $[\hat a^3]$, allows one to
obtain the following splitting scheme of the linearized system
$\ddot{\vec{\delta}}=\mathrm{J}(t)\cdot\vec{\delta}$
\begin{equation}
\begin{array}{lll}
  1:(\delta_1);&1:(\delta_6);&2:(\delta_2,\delta_3);\\
  2:(\delta_4,\delta_5);&3:(\delta_7,\delta_9,\delta_{11});
  &3:(\delta_8,\delta_{10},\delta_{12}).
\end{array}
 \label{eq80}
\end{equation}
Here we present the dimension of each independent subsystem (before
the colon) and the list of its variables (after the colon). From the
scheme (\ref{eq80}), one can see that two of four three-dimensional
subsystems corresponding to the splitting provided by group $[\hat
a^3]$ (see Eqs.~(\ref{eqk60})) in the case of the supergroup $[\hat
a^3,\hat\i]$ are decomposed into new independent subsystems of
dimensions equal to $1$ and $2$. Below we explain how one can obtain
the splitting schemes analogous to (\ref{eq80}).

\subsubsection{General case\label{Sec4.5.2}}

Now we consider the application of the Wigner theorem in case of an
arbitrary bush B$[G]$.

Let us consider a matrix $\mathrm{H}$ commuting with all matrices of
a reducible representation $\Gamma$ of the group $G$ that can be
decomposed into the irreps $\Gamma_j$ of this group as follows
\begin{equation}
\Gamma={\sum_j}^{\bigoplus} m_j\Gamma_j.
 \label{eq82}
\end{equation}
According to the Wigner theorem, the matrix $\mathrm{H}$ can be
reduced to a block-diagonal form with the blocks $\mathrm{D}_j$ of
dimensions $n_j\cdot m_j$, corresponding to the each $\Gamma_j$,
with $n_j$ being the dimension of the irrep $\Gamma_j$ entering
$m_j$ times into the decomposition (\ref{eq82}) of the
representation $\Gamma$ into the irreducible parts.

Moreover, each block $\mathrm{D}_j$ possesses a \emph{very specific}
form, namely, it consists of subblocks representing matrices
proportional to the \emph{identity matrix} $\mathrm{I}_{n_j}$ of the
dimension $n_j$ which repeat $m_j$ times along the rows and columns
of the block $\mathrm{D}_j$. We can illustrate the structure of a
certain block $\mathrm{D}_j=\mathrm{D}$ characterized by the numbers
$n_j=n$, $m_j=m$ as follows
\begin{equation}
\mathrm{D}=\left(
\begin{array}{cccc}
\mu_{11}\mathrm{I}_n& \mu_{12}\mathrm{I}_n&
\dots&\mu_{1m}\mathrm{I}_n \\
\mu_{21}\mathrm{I}_n& \mu_{22}\mathrm{I}_n&
\dots&\mu_{2m}\mathrm{I}_n \\
\dots&\dots&\dots&\dots\\
\mu_{m1}\mathrm{I}_n& \mu_{m2}\mathrm{I}_n&
\dots&\mu_{mm}\mathrm{I}_n
\end{array}\right),
 \label{eq84}
\end{equation}
where $\mathrm{I}_n$ is the $n\times n$ identity matrix.

In our case, the matrix $\mathrm{H}$ is the Jacobi matrix
$\mathrm{J}(t)$ of the linearized system
$\ddot{\vec{\delta}}=\mathrm{J}(t)\cdot\vec{\delta}$, $G$ is the
symmetry group of a given bush B$[G]$, $\Gamma$ is the mechanical
representation of this group. Each block $\mathrm{D}_j$ generates an
independent subsystem with $n_j\cdot m_j$ equations in the
decomposition of the linearized system. However, each of these
subsystems automatically splits into $n_j$ new subsystems consisting
of $m_j$ differential equations, as a consequence of the specific
structure of the block $\mathrm{D}_j$ (see Eq.~(\ref{eq84})).
Indeed, for example, if a certain $\mathrm{D}$-block for the matrix
$\mathrm{J}(t)$ possesses the form ($n_j=3$, $m_j=2$):
 \[
\mathrm{J}_j(t)=\left(
\begin{array}{cc}
\mu_{11}\mathrm{I}_3& \mu_{12}\mathrm{I}_3\\
\mu_{21}\mathrm{I}_3& \mu_{22}\mathrm{I}_3
\end{array}\right),
 \]
it is easy to check that we obtain the following three independent
pairs of the equations from the system
$\ddot{\vec{\delta}}=\mathrm{J}(t)\cdot\vec{\delta}$:
\begin{equation}
\begin{array}{l}
 \left\{
 \begin{array}{l}
\ddot{\delta}_1=\mu_{11}\delta_1+\mu_{12}\delta_4,\\
\ddot{\delta}_4=\mu_{21}\delta_1+\mu_{22}\delta_4,
 \end{array}
 \right.\\ \\
  \left\{
 \begin{array}{l}
\ddot{\delta}_2=\mu_{11}\delta_2+\mu_{12}\delta_5,\\
\ddot{\delta}_5=\mu_{21}\delta_2+\mu_{22}\delta_5,
 \end{array}
 \right.\\ \\
 \left\{
 \begin{array}{l}
\ddot{\delta}_3=\mu_{11}\delta_3+\mu_{12}\delta_6,\\
\ddot{\delta}_6=\mu_{21}\delta_3+\mu_{22}\delta_6.
 \end{array}
 \right.
\end{array}
 \label{eq85}
\end{equation}
Note that the dimension of each subsystem (\ref{eq85}) is equal to
$m_j=2$, while the total number of these subsystems is equal to
$n_j=3$.

\subsubsection{Irreducible representations and their basis vectors
for the dihedral group\label{Sec4.5.3}}

Hereafter, for simplicity, we will discuss only chains with an
\emph{even} number ($N$) of particles and illustrate the main ideas
with the example $N=12$.

The symmetry of an $N$-particle (monoatomic) chain is completely
described by the \emph{dihedral} group $G_0=D_N$ which can be
written as the union of two cosets with respect to its translational
subgroup $T_N=\{\hat e,\hat a,\hat a^2,\dots,\hat a^{N-1}\}$:
\begin{equation}
D_N=T_N\oplus T_N\cdot\hat\i.
 \label{eq90}
\end{equation}
Here $\hat\i$ is the inversion relative to the center of the chain.
The group $D_N$ is a non-Abelian group, since some of its elements
do not commute with each other (for example, $\hat\i\hat a=\hat
a^{-1}\hat\i$). As a consequence, the number of classes of conjugate
elements of this group is less than the total number ($2N$) of its
elements and some irreps $\Gamma_j$ are not one-dimensional. The
irreps of the dihedral group $D_N$ can be obtained by the well-known
induction procedure from those of its subgroup $T_N$. It turns out
that for $D_N$ with \emph{even} $N$ there are four one-dimensional
irreps, while all the other ($\frac{N}{2}-1$) irreps are
two-dimensional. We discussed the construction of these irreps in
\cite{FPU1}, where the following results were obtained.

Every irrep can be determined by two matrices $\mathrm{M}_j(\hat
a)$, $\mathrm{M}_j(\hat \i)$ corresponding to its generators $\hat
a$ and $\hat\i$, where $j$ is the number of this irrep. Four
one-dimension irreps ($j=1,2,3,4$) are real and are determined by
matrices \footnote{All combinations of signs are allowed in
(\ref{eq8080}).}
\begin{equation}
\mathrm{M}_j(\hat a)=\pm1,\qquad\mathrm{M}_j(\hat \i)=\pm1.
 \label{eq8080}
\end{equation}
All other irreps are two-dimensional and are determined by matrices
\[
\mathrm{M}_j(\hat a)=\left(
\begin{array}{cc}
\mu_j& 0\\
 0& \bar{\mu}_j
\end{array}\right),\qquad\mathrm{M}_j(\hat \i)=\left(
\begin{array}{cc}
 0& 1\\
 1&0
\end{array}\right),
\]
with $\mu_j=e^{\frac{2\pi i j}{N}}$, $\bar{\mu}_j=e^{-\frac{2\pi i
j}{N}}$ ($j\neq0,\, N/2$) \footnote{For the values $j=0$ and
$j=N/2$, two-dimensional representations turn out to be reducible
and they decompose into two pairs of one-dimensional irreps listed
in (\ref{eq8080}).}.

Let us find the basis vectors of the irreducible representations of
the dihedral group $[\hat a^m,\hat\i]$ for the case $m=3$ which
corresponds to the bush B$[\hat a^m,\hat\i]$. Let $\vec{\phi}$ and
$\vec{\psi}$ be the basis vectors of the two-dimensional invariant
subspace corresponding to the irrep with the matrix $\mathrm{M}(\hat
g)=\left(
\begin{array}{cc}
\gamma& 0\\
 0& \bar{\gamma}
\end{array}\right)$, where $\hat g=\hat a^m$ is the translational
generator of the dihedral group. They can be obtained from the
equations $\hat g\vec{\phi}=\gamma\vec{\phi}$ and $\hat
g\vec{\psi}=\bar{\gamma}\vec{\psi}$, respectively. For example,
using Eq.\,(\ref{eq212}) for the case $m=3$, we find
$\vec{\phi}=\{x,y,z~|~\gamma^{-1}x,\gamma^{-1}y,\gamma^{-1}z~|~
\gamma^{-2}x,\gamma^{-2}y,\gamma^{-2}z~|~\dots\}$,
$\vec{\psi}=\{\tilde{x},\tilde{y},\tilde{z}~|~\bar{\gamma}^{-1}\tilde{x},
\bar{\gamma}^{-1}\tilde{y},\bar{\gamma}^{-1}\tilde{z}~|~
\bar{\gamma}^{-2}\tilde{x},\bar{\gamma}^{-2}\tilde{y},
\bar{\gamma}^{-2}\tilde{z}~|~\dots\}$. Here ($x,y,z$) and
($\tilde{x},\tilde{y},\tilde{z}$) are arbitrary constants which
these vectors depend on.

Taking into account the presence of the matrix $\mathrm{M}(\hat \i)=
\left(\begin{array}{cc} 0& 1\\1&0\end{array}\right)$ in every
two-dimensional irrep, one can state that
\begin{equation}
\hat\i\vec{\phi}=\vec{\psi},\qquad\hat\i\vec{\psi}=\vec{\phi}.
 \label{eq9090}
\end{equation}
Because of these relations, there appear certain connections between
the arbitrary constants ($x,y,z$) and
($\tilde{x},\tilde{y},\tilde{z}$). As a consequence, the basis
vectors $\vec{\phi}$ and $\vec{\psi}$, for each two-dimensional
irrep of the group $[\hat a^3,\hat\i]$, depends on only three
arbitrary parameters: $x$, $y$ and $z$. In turn, this means that
each two-dimensional irrep enters exactly \emph{three} times into
the decomposition of the mechanical representation of the dihedral
group $[\hat a^3,\hat\i]$.

Unlike this, the one-dimensional irreps of the dihedral group $[\hat
a^3,\hat\i]$ are contained in the mechanical representation less
than $3$ times. Indeed, let us consider the basis vectors
$\vec{\phi}$ and $\vec{\psi}$ of the one-dimensional irreps of the
group $[\hat a^3]$ determined by the matrices $\mathrm M(a^3)=1$ and
$\mathrm M(a^3)=-1$, respectively, for the case $m=3$, $N=12$:
 \[
\begin{array}{l}
\vec{\phi}=\{x,y,z~|~x,y,z~|~x,y,z~|~x,y,z\},\\
\vec{\psi}=\{x,y,z~|~-x,-y,-z~|~x,y,z~|~-x,-y,-z\}.
\end{array}
 \]
These vectors can be obtained from Eq.~(\ref{eq212}) by letting
$\gamma=1$ and $\gamma=-1$. If the vector $\vec{\phi}$ is not only
the basis vector of the irrep of the group $[\hat a^3]$, but also is
the basis vector of a certain one-dimensional irrep of the dihedral
group $[\hat a^3,\hat\i]$, it must satisfy the equations
$\hat\i\vec{\phi}=\vec{\phi}$, for the irrep $\Gamma_1$ ($\mathrm
M(\hat\i)=1$) and $\hat\i\vec{\phi}=-\vec{\phi}$, for the irrep
$\Gamma_2$ ($\mathrm M(\hat\i)=-1$). We obtain $z=-x$, $y=0$ from
the former equation, and $z=x$ for the latter equation. Thus
 \[
\begin{array}{l}
\vec{\phi}[\Gamma_1]=\{x,0,-x~|~x,0,-x~|~x,0,-x~|~x,0,-x\},\\
\vec{\phi}[\Gamma_2]=\{x,y,x~|~x,y,x~|~x,y,x~|~x,y,x\}.
\end{array}
 \]

In the same manner, we obtain the basis vectors
$\vec{\psi}[\Gamma_3]$ and $\vec{\psi}[\Gamma_4]$ from the equations
$\hat\i\vec{\psi}=\vec{\psi}$ and $\hat\i\vec{\psi}=-\vec{\psi}$,
respectively:
 \[
\begin{array}{l}
\vec{\psi}[\Gamma_3]=\{x,y,x~|~-x,-y,-x~|~x,y,x~|~-x,-y,-x\},\\
\vec{\psi}[\Gamma_4]=\{x,0,-x~|~-x,0,x~|~x,0,-x~|~-x,0,x\}.
\end{array}
 \]
From the above results, we conclude that the irreps $\Gamma_1$ and
$\Gamma_4$ are contained once, while the irreps $\Gamma_2$ and
$\Gamma_3$ are contained twice in the decomposition of the
mechanical representation for the considered chain.

The generalization of these results to the case of the dihedral
group $[\hat a^m,\hat\i]$ with \emph{arbitrary} $m$ is trivial.

The splitting scheme (\ref{eq80}) for the bush B$[\hat a^3,\hat\i]$
for the monoatomic chain with $N=12$ particles can be now explained
as follows. There are five irreps ($\Gamma_1$, $\Gamma_2$,
$\Gamma_3$, $\Gamma_4$, $\Gamma_5$) of the group $[\hat
a^3,\hat\i]\equiv\{\hat e,\hat a^3,\hat a^6,\hat
a^9~|~\hat\i,\hat\i\hat a^3,\hat\i\hat a^6,\hat\i\hat a^9\}\equiv
D_4$. As have just shown, the one-dimensional irreps $\Gamma_1$
($n_1=1$) and $\Gamma_4$ ($n_4=1$) are contained once ($m_1=1$,
$m_4=1$) in the decomposition of the mechanical representation
$\Gamma$ for our chain. On the other hand, the one-dimensional
irreps $\Gamma_2$ ($n_2=1$) and $\Gamma_3$ ($n_3=1$) are contained
twice ($m_2=2$, $m_3=2$) in $\Gamma$, while two-dimensional irrep
$\Gamma_5$ ($n_5=2$) is contained thrice ($m_5=3$) in $\Gamma$. The
twelve variables $\delta_j$ from Eq.~(\ref{eq80}) are associated
with the irreps of the group $D_4$ in the following manner:
\begin{equation}
\begin{array}{lll}
  \delta_1\rightarrow\Gamma_1(1);&&\\
  \delta_2\rightarrow\Gamma_2(1),&\delta_3\rightarrow\Gamma_2(2);&\\
  \delta_4\rightarrow\Gamma_3(1),&\delta_5\rightarrow\Gamma_3(2);&\\
  \delta_6\rightarrow\Gamma_4(1);&&\\
  (\delta_7,\delta_8)\rightarrow\Gamma_5(1),&
  (\delta_9,\delta_{10})\rightarrow\Gamma_5(2),&
  (\delta_{11},\delta_{12})\rightarrow\Gamma_5(3).
\end{array}
 \label{eq500}
\end{equation}
Here, in parenthesis, we give the index of the copy of the irrep
$\Gamma_i$ (whose dimension is equal to $n_i$) in the decomposition
of the mechanical representation $\Gamma$. Note that the total
number of such copies determines how many times $m_i$ the irrep
$\Gamma_i$ is contained in $\Gamma$. On the other hand, as we
already know, $m_i$ shows us the dimension of the subsystems $L_j$,
while $n_i$ determines the total number of $L_j$ with the same
dimension associated with $\Gamma_i$. As a result, we obtain the
splitting scheme (\ref{eq80}).

The above discussed decomposition of the full linearized system
$\ddot{\vec{\delta}}=\mathrm{J}(t)\cdot\vec{\delta}$ into
independent subsystems $L_j$ of small dimensions permits one to
analyze efficiently the stability of a given bush in the monoatomic
chain with arbitrary large number of particles ($N$).

\begin{figure}
\centering \setlength{\unitlength}{1mm}
\begin{picture}(130,45)(0,0)
\put(5,5){\includegraphics[width=120mm,height=36mm]{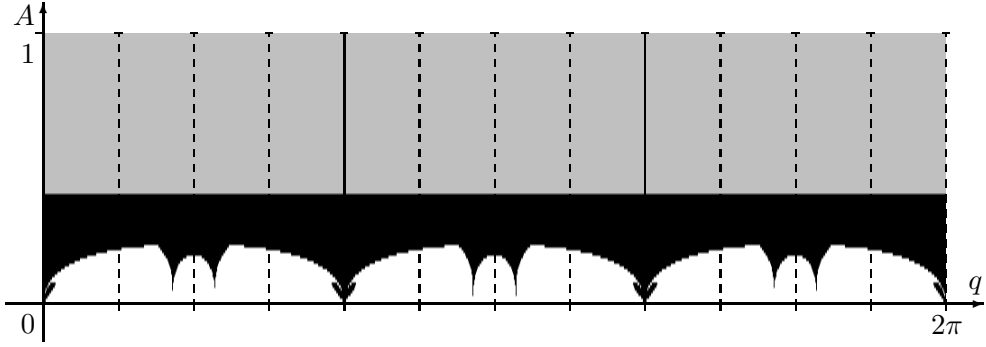}}
\drawgridA
\put(45,5){\line(0,1){36}}
\put(85,5){\line(0,1){36}}
\end{picture}
\caption{\label{fig1} Regions of stability (white color) of
different modes of the FPU-$\alpha$ chain, interacting
parametrically with the one-dimensional bush B$[a^3,i]$.}
\end{figure}

Using this idea, the stability diagrams for all the one-dimensional
bushes in both FPU-$\alpha$ and FPU-$\beta$ chains were obtained in
\cite{FPU2}. As an example, in Fig.~\ref{fig1}, we reproduce the
stability diagram for the bush B$[\hat a^3,\hat\i]$ for the
FPU-$\alpha$ chain from that paper. In this diagram, each point
$(A,q)$ determines a certain value of the bush mode amplitude $A$
and a certain value of the wave number $q=\frac{2\pi j}{N}$ that is
associated with the index $j$ of a fixed mode. The black points
$(A,q)$ correspond to the case where the mode $j=q\frac{N}{2\pi}$
becomes excited because of its parametric interaction with the mode
of the bush B$[\hat a^3,\hat\i]$. The white color denotes the
opposite case: the corresponding mode $j$, being zero at the initial
instant, continues to be zero in spite of its interaction with the
considered bush. Such a diagram allows one to study stability of
one-dimensional bushes not only for finite $N$, but also for the
case $N\rightarrow\infty$ (some more details can be found in
\cite{FPU2}).

\section{Some additional examples \label{Sec5}}

\subsection{Stability of the $\pi$-mode in the FPU-$\alpha$
chain\label{Sec5.1}}

The symmetry group of the $\pi$-mode
$\vec{X}(t)=\{A(t),-A(t),A(t),-A(t),A(t),-A(t),\dots\}$ in the
FPU-$\alpha$ chain is $[\hat a^2,\hat\i]$. Using results of the
previous sections, one can deduce for $N$ divisible by $4$ that the
linearized system in the vicinity of this mode splits into four
individual equations and a number of two-dimensional systems of
differential equations \footnote{If $N$ is an even number, but $N
\mod 4\neq0$, one obtains two equations of harmonic oscillators and
a number of pairs of coupled equations.}. The first pair of
individual equations represents two independent harmonic
oscillators, while the second pair represents two Mathieu equations.
All the other systems represent pairs of coupled equations:
\begin{equation}
\begin{array}{l}
\ddot{\delta}_j+4\sin^2\left(\frac{\pi
j}{N}\right)\delta_j=\eta\sin\left(\frac{2\pi j}{N}\right)
\delta_{\frac{N}{2}-j}\cos(2t),\\
\ddot{\delta}_{\frac{N}{2}-j}+4\cos^2\left(\frac{\pi
j}{N}\right)\delta_{\frac{N}{2}-j}=\eta\sin\left(\frac{2\pi
j}{N}\right)\delta_j\cos(2t),
 \label{eq5100}
\end{array}
\end{equation}
where $\eta=\frac{8A}{\sqrt{N}}$, $j=1,2,\dots,\frac{N}{4}-1$. In
(\ref{eq5100}), $A$ is the amplitude of the $\pi$-mode, i.e.
$A=\max|A(t)|$. Eqs.~(\ref{eq5100}) can be rewritten as follows:
\begin{equation}
\begin{array}{l}
\ddot{x}+4\sin^2\left(\frac{q}{2}\right)x=\eta\sin(q)y\cos(2t),\\
\ddot{y}+4\cos^2\left(\frac{q}{2}\right)y=\eta\sin(q)x\cos(2t),
 \label{eq5101}
\end{array}
\end{equation}
where $q=\frac{2\pi j}{N}$ is the wave number, $x(t)=\delta_j(t)$,
$y(t)=\delta_{\frac{N}{2}-j}(t)$. These results were obtained and
discussed in \cite{FPU1} using a different (in comparison with the
present paper) method. Indeed, there we obtained \emph{exact}
equations for the FPU-$\alpha$ chain in the \emph{modal} space and
only then linearized them near the $\pi$-mode.

As it was discussed in \cite{FPU1}, the system (\ref{eq5101}) turns
out to be rather remarkable: the $\pi$-mode (or one-dimensional bush
B$[\hat a^2,\hat\i]$) loses its stability \emph{simultaneously} with
respect to interaction with all the other modes. In other words, the
threshold $\eta_c$ for the loss of stability for the bush B$[\hat
a^2,\hat\i]$ turns out to be \emph{the same} for all the values of
$q$, i.e. for all the subsystems (\ref{eq5100}):
\begin{equation}
\eta_c=2.42332\dots
 \label{eq5102}
\end{equation}
This property of the linearized system (\ref{eq5101}) was discussed
in more detail in \cite{FPU1}. The stability diagram for the bush
B$[\hat a^2,\hat\i]$ can be found in \cite{FPU2}.

\subsection{Stability of the $\pi$-mode in the FPU-$\beta$
chain\label{Sec5.2}}

This case is of a particular interest since the linearized system
for the $\pi$-mode in the FPU-$\beta$ chain possesses a \emph{higher
symmetry} compared to that in the FPU-$\alpha$ chain. Indeed, for
the FPU-$\beta$ model, the interparticle potential is an \emph{even}
function, and some additional symmetry elements of the dynamical
equations appear as a consequence of this fact.

Let us introduce an operator $\hat u$ that changes signs of the
displacements of all the particles:
\begin{equation}
\hat u\{x_1,x_2,\dots,x_N\}=\{-x_1,-x_2,\dots,-x_N\}.
 \label{eq5290}
\end{equation}
This operator leaves the FPU-$\beta$ Hamiltonian unchanged because
of its evenness.

Therefore, the symmetry group of the FPU-$\beta$ dynamical equations
turns out to be a \emph{supergroup} $G'_0$ with respect to the
symmetry group $G_0=[\hat a,\hat\i]$ of the FPU-$\alpha$ chain:
\begin{equation}
G'_0=G_0\oplus G_0\cdot\hat u
 \label{eq5291}
\end{equation}
(note that $\hat u^2=\hat e$).

As a result, one can classify bushes of modes in the FPU-$\beta$
chain by subgroups of the group $G'_0$ rather than by subgroups of
the $G_0$. The invariant manifolds (bushes of modes) in the
FPU-$\beta$ model with respect to the group $G'_0$ were found by
Rink in \cite{BR}. Some additional details of this problem were
discussed in our paper \cite{FPU2} (in particular, the dynamical
equations and stability of these bushes of modes).

Considering  $G'_0$ from (\ref{eq5291}) as the parent group, we
discover that the $\pi$-mode in the FPU-$\beta$ chain must be
characterized by the group $G=[\hat a^2,\hat\i,\hat a\hat u]$,
unlike the group $[\hat a^2,\hat\i]$ characterizing the $\pi$-mode
in the FPU-$\alpha$ chain with the parent group $[\hat a,\hat\i]$.

The third generator \footnote{Obviously, the operators $\hat a$ and
$\hat u$ commute with each other.} $\hat a\hat u\equiv\hat u\hat a$
of the group $G=[\hat a^2,\hat\i,\hat a\hat u]$ acts on the
configuration vector $\vec{X}$ as follows:
\begin{equation}
\hat a\hat u\vec{X}=\{-x_N,-x_1,-x_2,\dots,-x_{N-1}\}.
 \label{eq5292}
\end{equation}

Above, we specified the group $G$ by \emph{three} generators, but it
can be determined by only \emph{two} generators. Indeed, if
\begin{equation}
\hat p=\hat a\hat u,
 \label{eq5293}
\end{equation}
then $\hat p^2=\hat a\hat u\hat a\hat u=\hat a^2\hat u^2=\hat a^2$.
Thus, the first generator in the list $G=[\hat a^2,\hat\i,\hat a\hat
u]$ is simply the square of the third generator $\hat p=\hat a\hat
u$.

For simplicity, let us consider the case $N=6$. Then $[\hat
a^2]=\{\hat e,\hat a^2,\hat a^4\}$, ($\hat a^6=\hat e$) is a cyclic
group of the order $3$ and the full group $G$ contains
$3\cdot2\cdot2=12$ elements. With the aid of the operators $\hat
p=\hat a\hat u$ and  $\hat q=\hat\i$, we can obtain all the elements
of the group
\begin{equation}
G=[\hat a^2,\hat\i,\hat a\hat u]\equiv[\hat p,\hat q]
 \label{eq5295}
\end{equation}
as follows:
\begin{equation}
G=T_6\oplus T_6\cdot\hat q,
 \label{eq5296}
\end{equation}
where $T_6$ is the cyclic group of the order $6$:
\begin{equation}
T_6=\{\hat e,\hat p,\hat p^2,\hat p^3,\hat p^4,\hat p^5\}.
 \label{eq5297}
\end{equation}

The following generating relations, fully determining the group
$G=[\hat p,\hat q]$, can be obtained:
\begin{equation}
\hat p^6=\hat e,\qquad\hat q^2=\hat e,\qquad\hat q\hat p=\hat
p^5\hat q.
 \label{eq5298}
\end{equation}
From these relations, one can see that $G$ is the \emph{dihedral}
group $D_6$ (by the way, it is isomorphic to the point groups
$C_{6v}$, $D_{3h}$ and $D_{3d}$, as well).

The irreducible representations of the dihedral group were discussed
in \cite{FPU1} (see also Sec.\,\ref{Sec4.5} of the present paper).
There are four one-dimensional and $\left(\frac{N}{2}-1\right)$
two-dimensional irreps of the group $D_N$ with an even index $N$.
For simplicity, let us discuss the case $N=6$ only (the
generalization to arbitrary values of $N$ turns out to be trivial).

All irreps of the group $D_6$ can be constructed from the irreps of
its subgroup $T_6$ (see (\ref{eq5296})) with the aid of the
induction procedure. As a result, we obtain the following irreps of
the group $G=[\hat p,\hat q]$ presented in Table\,\ref{table3}. In
this table, each irrep $\Gamma_j$ ($j=1,2,\dots,6$) is determined by
the matrices $\mathrm{M}(\hat p)$ and $\mathrm{M}(\hat q)$,
corresponding to the generators $\hat p$ and $\hat q$ of the group
$D_6$.
\begin{table}
\caption{Irreducible representations of the dihedral group
$D_6\equiv[\hat p,\hat q]$. In this table $\mu=e^{\frac{\pi i}{3}}$,
$\nu=\mu^2=e^{\frac{2\pi i}{3}}$, $\bar{\mu}$ and $\bar{\nu}$ are
conjugate complex numbers with respect to $\mu$ and $\nu$,
respectively.\label{table3}}
\begin{ruledtabular}
\begin{tabular}{c|cc}
 &$\mathrm{M}(\hat p)$&$\mathrm{M}(\hat q)$\\
\hline
 $\Gamma_1$&   $1$&   $1$\\
 $\Gamma_2$&   $1$&  $-1$\\
 $\Gamma_3$&  $-1$&   $1$\\
 $\Gamma_4$&  $-1$&  $-1$\\
 $\Gamma_5$&
 $\left(
\begin{array}{cc}
\mu& 0\\
 0& \bar{\mu}
\end{array}\right)$
&
 $\left(
\begin{array}{cc}
0& 1\\
 1& 0
\end{array}\right)$\\
 $\Gamma_6$&
 $\left(
\begin{array}{cc}
\nu& 0\\
 0& \bar{\nu}
\end{array}\right)$
&
 $\left(
\begin{array}{cc}
0& 1\\
 1& 0
\end{array}\right)$
\end{tabular}
\end{ruledtabular}
\end{table}

All the invariant subspaces of the configuration space of the chain
with $N=6$ particles, corresponding to these irreps, and their basis
vectors can be obtained as follows. Let us find the basis vector
$\vec{\phi}$ of a certain one-dimensional irrep $\Gamma$ determined
by the matrices $\mathrm{M}(\hat p)=\gamma$ and $\mathrm{M}(\hat
q)=\delta$ from Table\,\ref{table3} ($\gamma=\pm1$, $\delta=\pm1$).
Thus, the vector $\vec{\phi}$ must satisfy the relations:
\begin{subequations}
\label{eq5200}
\begin{eqnarray}
&&\hat p\vec{\phi}=\gamma\vec{\phi},\label{eq5200a}\\
&&\hat q\vec{\phi}=\delta\vec{\phi}.\label{eq5200b}
\end{eqnarray}
\end{subequations}
 From (\ref{eq5200a}), we obtain
\begin{equation}
\vec{\phi}=\{x,-\gamma^5x,\gamma^4x,-\gamma^3x,\gamma^2x,-\gamma
x\},
 \label{eq5201}
\end{equation}
where $x$ is an arbitrary constant. For $\gamma=1$, the basis vector
$\vec{\phi}$ from (\ref{eq5201}) transforms to
\begin{equation}
\vec{\phi}_1=\{x,-x,x,-x,x,-x\},
 \label{eq5202}
\end{equation}
while for $\gamma=-1$ it transforms to
\begin{equation}
\vec{\phi}_2=\{x,x,x,x,x,x\}.
 \label{eq5203}
\end{equation}
Now we must also demand (\ref{eq5200b}) to hold. Remembering that
$\hat q=\hat\i$, and, therefore, that $\hat
q\vec{\phi}=\{-x_6,-x_5,-x_4,-x_3,-x_2,-x_1\}$, we obtain
 \[
 \begin{array}{l}
\hat q\vec{\phi}_1=\{x,-x,x,-x,x,-x\}\equiv(1)\vec{\phi}_1,\\
\hat q\vec{\phi}_2=\{-x,-x,-x,-x,-x,-x\}\equiv(-1)\vec{\phi}_2.
 \end{array}
 \]
 This means that $\vec{\phi}_1$ from (\ref{eq5202}) is the basis
vector of the irrep $\Gamma_1$ ($\gamma=1$, $\delta=1$, see
Eqs.~(\ref{eq5200})), while $\vec{\phi}_2$ from (\ref{eq5203}) is
the basis vector of the irrep $\Gamma_4$ ($\gamma=-1$, $\delta=-1$).
Thus, each irrep $\Gamma_1$ and $\Gamma_4$ is contained once in the
decomposition of the mechanical representations.

On the other hand, the basis vectors of irreps $\Gamma_2$
($\gamma=1$, $\delta=-1$) and $\Gamma_3$ ($\gamma=-1$, $\delta=1$)
are equal to zero. Indeed, demanding $\hat
q\vec{\phi}_1=(-1)\vec{\phi}_1$ ($\delta=-1$) one concludes that
$x=-x$ and, therefore, $x=0$ and $\vec{\phi}_1\equiv0$. The same
result originates also from the equation $\hat
q\vec{\phi}_2=(1)\vec{\phi}_2$ ($\delta=1$), namely,
$\vec{\phi}_2\equiv0$. Thus, we must conclude that the irreps
$\Gamma_2$ and $\Gamma_3$ \emph{are not contained} in the
decomposition of the mechanical representation of our chain into
irreducible representations (the corresponding invariant subspaces
turn out to be null spaces).

There are two basis vectors, $\vec{\phi}$ and $\vec{\psi}$, for each
two-dimensional irreps from Table\,\ref{table3}. For the irrep
$\Gamma_5$ the following relations must hold
\begin{subequations}
\label{eq5205}
\begin{eqnarray}
&&\hat p\vec{\phi}=\mu\vec{\phi},\label{eq5205a}\\
&&\hat p\vec{\psi}=\bar{\mu}\vec{\psi},\label{eq5205b}\\
&&\hat q\vec{\phi}=\vec{\psi},\label{eq5205c}\\
&&\hat q\vec{\psi}=\vec{\phi}\label{eq5205d}.
\end{eqnarray}
\end{subequations}
 Comparing (\ref{eq5205a}) and (\ref{eq5205b}) with (\ref{eq5200a}),
we can write
\begin{equation}
\vec{\phi}=\{x,-\mu^5x,\mu^4x,-\mu^3x,\mu^2x,-\mu x\},
 \label{eq5206}
\end{equation}
\begin{equation}
\vec{\psi}=\{y,-\bar{\mu}^5y,\bar{\mu}^4y,-\bar{\mu}^3y,
\bar{\mu}^2y,-\bar{\mu}y\},
 \label{eq5207}
\end{equation}
where $x$ and $y$ are two \emph{different} constants.

Then, from (\ref{eq5205c}) or (\ref{eq5205d}), we obtain a certain
relations between the constants $x$ and $y$:
\begin{equation}
y=\mu x.
 \label{eq300}
\end{equation}
Substituting this value of $y$ into (\ref{eq5207}) we, finally,
conclude that the irrep $\Gamma_5$ is contained only once in the
decomposition of the mechanical representation of our FPU-$\beta$
chain. Indeed, the basis vectors $\vec{\phi}$, $\vec{\psi}$ turn out
to depend on only one constant $x$ which can be determined from the
normalization condition. The same conclusion is valid also for the
second irrep $\Gamma_6$ of the group $D_6$.

The generalization of this conclusion to the case of the FPU-$\beta$
chain with an arbitrary even number $N=2n$ of particles can be
achieved trivially. Namely, each two-dimensional irrep of the group
$D_{2n}$ is contained only \emph{once} in the mechanical
representation of this group for the FPU-$\beta$ chain.
One-dimensional irreps are contained in the mechanical
representation once or not at all. Remembering the above discussion
about the application of the Wigner theorem to splitting the
linearized system
$\ddot{\vec{\delta}}=\mathrm{J}(t)\cdot\vec{\delta}$, we discover
that for the FPU-$\beta$ chain (unlike FPU-$\alpha$ chain) this
system is decomposed into individual differential equations of the
second order, i.e. we have the \emph{complete splitting} in this
case.

The following remark should be done to avoid a possible
misunderstanding. The form of the $\pi$-mode
\begin{equation}
\vec{X}(t)=\{A(t),-A(t),A(t),-A(t),A(t),-A(t)\},
 \label{eq350}
\end{equation}
is one and the same \footnote{Note that $A(t)$ in Eq.~(\ref{eq350})
are \emph{different} functions of time for the FPU-$\alpha$ and
FPU-$\beta$ chains (see below).} for both FPU-$\alpha$ and
FPU-$\beta$ chains. Therefore, its symmetry group can be written as
$[\hat a^2,\hat\i,\hat a\hat u]$ rather than $[\hat a^2,\hat\i]$
since the operator $\hat a\hat u$ does not change the pattern
(\ref{eq350}) not only for the FPU-$\beta$ chain, but also for the
FPU-$\alpha$ chain.

Nevertheless, the linearized system
$\ddot{\vec{\delta}}=\mathrm{J}(t)\cdot\vec{\delta}$, for the
FPU-$\beta$ chain, is invariant with respect to the operator $\hat
a\hat u$, while that for the FPU-$\alpha$ chain is not invariant
under the action of $\hat a\hat u$. Indeed, the linearized (near the
$\pi$-mode) system for the FPU-$\alpha$ chain reads:
\begin{equation}
\begin{array}{l}
\ddot{\delta}_1=[\delta_2-2\delta_1+\delta_6]-4A(t)\cdot[\delta_2-\delta_6],\\
\ddot{\delta}_2=[\delta_3-2\delta_2+\delta_1]+4A(t)\cdot[\delta_3-\delta_1],\\
\ddot{\delta}_3=[\delta_4-2\delta_3+\delta_2]-4A(t)\cdot[\delta_4-\delta_2],\\
\ddot{\delta}_4=[\delta_5-2\delta_4+\delta_3]+4A(t)\cdot[\delta_5-\delta_3],\\
\ddot{\delta}_5=[\delta_6-2\delta_5+\delta_4]-4A(t)\cdot[\delta_6-\delta_4],\\
\ddot{\delta}_6=[\delta_1-2\delta_6+\delta_5]+4A(t)\cdot[\delta_1-\delta_5].
 \label{eq351}
\end{array}
\end{equation}
On the other hand, for the FPU-$\beta$ chain, the linearized system
reads:
\begin{equation}
\begin{array}{l}
\ddot{\delta}_1=[\delta_2-2\delta_1+\delta_6]\cdot[1+12A^2(t)],\\
\ddot{\delta}_2=[\delta_3-2\delta_2+\delta_1]\cdot[1+12A^2(t)],\\
\ddot{\delta}_3=[\delta_4-2\delta_3+\delta_2]\cdot[1+12A^2(t)],\\
\ddot{\delta}_4=[\delta_5-2\delta_4+\delta_3]\cdot[1+12A^2(t)],\\
\ddot{\delta}_5=[\delta_6-2\delta_5+\delta_4]\cdot[1+12A^2(t)],\\
\ddot{\delta}_6=[\delta_1-2\delta_6+\delta_5]\cdot[1+12A^2(t)].
 \label{eq352}
\end{array}
\end{equation}

The operator $\hat a\hat u$ acts on the vector $\vec{\delta}$ as a
follows:
\begin{equation}
\hat a\hat u\vec{\delta}\equiv\hat a\hat
u\{\delta_1,\delta_2,\delta_3,\delta_4,\delta_5,\delta_6\}=\{-\delta_6,
-\delta_1,-\delta_2,-\delta_3,-\delta_4,-\delta_5\}.
 \label{eq355}
\end{equation}
The substitution
\begin{equation}
\begin{array}{l}
\delta_1\rightarrow-\delta_6,\quad\delta_2\rightarrow-\delta_1,
\quad\delta_3\rightarrow-\delta_2,\\
\delta_4\rightarrow-\delta_3,\quad\delta_5\rightarrow-\delta_4,
\quad\delta_6\rightarrow-\delta_5,
 \label{eq356}
\end{array}
\end{equation}
transforms, for example, the first equation of the system
(\ref{eq351}) to the form
\[
\ddot{\delta}_6=[\delta_1-2\delta_6+\delta_5]-4A(t)\cdot[\delta_1-\delta_5],
\]
which differs by the sign in front of $4A(t)$ from the sixth
equation of the system (\ref{eq351}). Contrariwise, the
transformation (\ref{eq356}) leaves the system (\ref{eq352}) for the
FPU-$\beta$ chain unchanged (it leads only to some transpositions of
the individual equation in (\ref{eq352})).

What does it mean? Each bush (the $\pi$-mode, in our case) is
associated with a certain subgroup of the parent group $G_0$, the
symmetry group of the original nonlinear \emph{dynamical equations}
of the considered mechanical system. The operator $\hat u$ does not
belong to the symmetry group $G_0$ of the FPU-$\alpha$ chain (unlike
the case of the FPU-$\beta$ chain!) and, therefore, there are no
subgroups of the group $G_0$ whose elements contain $\hat u$.
Precisely this fact can explain why we must not take into
consideration the operator $\hat a\hat u$ for the case of the
FPU-$\alpha$ chain, even through this operator does not change the
pattern (\ref{eq350}) of the $\pi$-mode.

The function $A(t)$ from the displacement pattern (\ref{eq350}) of
the $\pi$-mode is determined by the dynamical equation of the
one-dimensional bush B$[\hat a^2,\hat\i]$. As it was already
discussed, this equation can be obtained by substituting the vector
$\vec{X}(t)$ from Eq.~(\ref{eq350}) into the nonlinear dynamical
equations of the FPU chain. For the FPU-$\alpha$ model this equation
turns out to be the equation of a harmonic oscillator $\ddot
A+4A=0$, while for the FPU-$\beta$ model it is the Duffing equation
$\ddot A+4A+16A^3=0$.

Thus, in both cases, the exact expression for the function $A(t)$
can be found. For studying the stability of the $\pi$-mode, we must
substitute the corresponding expression for $A(t)$ into the
linearized system
$\ddot{\vec{\delta}}=\mathrm{J}(t)\cdot\vec{\delta}$. As we saw in
the present section, the linearized system splits into individual
equations for the FPU-$\beta$ chain, while for the FPU-$\alpha$
chain it can be decomposed into four individual equations and
$(N-4)/2$ pairs of differential equations. All these equations turn
out to be equations of the second order with time-periodic
coefficients depending on the function $A(t)$ \cite{FPU1,FPU2}.

Let us note that stability of the $\pi$-mode in the FPU-$\alpha$ and
FPU-$\beta$ chains was investigated by different methods in a large
number of papers (see, for example, [\onlinecite{Bud,Sand,Flach},\,
\onlinecite{FPU1,FPU2,PR,Yosh,Shin,CLL,AntiFPU}]), but, to our best
understanding, the influence of symmetry of these mechanical models
on stability analysis was not discussed. Unlike the above cited
works, the bush stability analysis presented in this paper based on
the symmetry-related arguments only. Therefore, our conclusion about
the difference in splitting scheme of linearized systems for the
$\pi$-mode in the FPU-$\alpha$ and FPU-$\beta$ chains can be
automatically extended to all the other nonlinear chains with the
same symmetry characteristics. In particular, we can conclude that
the splitting of the linearized system into individual equations can
be performed not only for the FPU-$\beta$ chain, but for every chain
with an \emph{even} potential of the interparticle interaction. In
contrast, it is impossible for the chains with \emph{arbitrary}
potential, and the FPU-$\alpha$ model is a simple illustration of
this proposition.

We would like to focus on the paper \cite{Yosh}, where some
analytical results were obtained for the stability of the $\pi$-mode
in the nonlinear chains with a \emph{general even} form of the
interparticle interaction potential. The author of \cite{Yosh}
succeeded in his analysis thanks to the decomposition of the
linearized system into individual equations, but such analysis
cannot be extended to the FPU-$\alpha$ chain precisely because the
potential of this model is \emph{odd}.

\subsection{Stability of two-dimensional bush B$[\hat a^4,\hat\i]$ in
the FPU-$\alpha$ chain\label{Sec5.3}}

Now, let us consider the stability of the two-dimensional bush
B$[\hat a^4,\hat\i]$ that can be determined by the displacement
pattern \cite{FPU2}
\begin{equation}
\vec{X}(t)=\{A(t),B(t),-B(t),-A(t)~|~A(t),B(t),-B(t),-A(t)~|~
A(t),B(t),-B(t),-A(t)\}
 \label{eq411}
\end{equation}
(for simplicity, we start with the case $N=12$).

The symmetry group of this bush is the dihedral group $D_3$ with
translational subgroup
\begin{equation}
T_3=\{\hat e,\hat a^4,\hat a^8\}\qquad(\hat a^{12}=\hat e).
 \label{eq415}
\end{equation}
This non-Abelian group ($\hat\i\hat a^4=\hat a^8\hat\i$) consists of
six elements determined by the equation
\begin{equation}
D_3=T_3\oplus T_3\cdot\hat\i,
 \label{eq416}
\end{equation}
and possesses the following three irreps presented in
Table~\ref{table10} (we give there the two-dimensional irrep not
only in the complex form, but in the real form also).
\begin{table}
\caption{Irreducible representations of the dihedral group $D_3$.
Here $\mu=e^{\frac{2\pi i}{3}}$, $\bar{\mu}=e^{-\frac{2\pi i}{3}}$,
$\hat p=\hat a^4$, $\hat q=\hat i$.\label{table10}}
\begin{ruledtabular}
\begin{tabular}{c|cc|cc}
 &\multicolumn{2}{c}{Complex form}\vline&\multicolumn{2}{c}{Real form}\\
\cline{1-5}
 &$\mathrm{M}(\hat p)$&$\mathrm{M}(\hat q)$&
 $\mathrm{M}(\hat p)$&$\mathrm{M}(\hat q)$\\
\hline
 $\Gamma_1$&   $1$&   $1$&  $1$&  $1$\\
 $\Gamma_2$&   $1$&  $-1$&  $1$& $-1$\\
 $\Gamma_3$&
 $\left(
\begin{array}{cc}
\mu& 0\\
 0& \bar{\mu}
\end{array}\right)$
&
 $\left(
\begin{array}{cc}
0& 1\\
 1& 0
\end{array}\right)$
&
 $\left(
\begin{array}{cc}
-\frac{1}{2}& \frac{\sqrt{3}}{2}\\
 -\frac{\sqrt{3}}{2}& -\frac{1}{2}
\end{array}\right)$
&
 $\left(
\begin{array}{rr}
1& 0\\
 0& -1
\end{array}\right)$
\end{tabular}
\end{ruledtabular}
\end{table}

The basis vectors of the one-dimensional irreps of the group $T_3$
(\ref{eq415}) can be found from the relation
\begin{equation}
\hat p\vec{\phi}=\gamma\vec{\phi},
 \label{eq400}
\end{equation}
where $\gamma=1,\mu,\bar{\mu}$. From this equation we obtain
\begin{eqnarray}
\hat
p\vec{\phi}=\{x_9,x_{10},x_{11},x_{12}~|~x_1,x_2,x_3,x_4~|~x_5,x_6,x_7,
x_8\}=\nonumber\\
\gamma\{x_1,x_2,x_3,x_4~|~x_5,x_6,x_7,x_8~|~x_9,x_{10},x_{11},x_{12}\}
 \label{eq401}
\end{eqnarray}
and then
\begin{equation}
\vec{\phi}=\{x_1,x_2,x_3,x_4~|~\bar{\gamma}x_1,\bar{\gamma}x_2,
\bar{\gamma}x_3,\bar{\gamma}x_4~|~\gamma x_1,\gamma x_2, \gamma
x_3,\gamma x_4\}.
 \label{eq402}
\end{equation}
This basis vector depends on four arbitrary constants ($x_1$, $x_2$,
$x_3$, $x_4$) and, therefore, it determines a four-dimensional
subspace invariant under the translational group $T_3$
(\ref{eq415}), associated with the irrep $\Gamma$ defined by the
one-dimensional matrix $\mathrm{M}(\hat p)=\gamma$.

The basis vectors of one-dimensional irreps of the whole group $D_3$
from (\ref{eq416}) can be obtained with the aid of the equation
$\hat\i\vec{\phi}=\vec{\phi}$ (for the irrep $\Gamma_1$) and with
the aid of the equation $\hat\i\vec{\phi}=-\vec{\phi}$ (for the
irrep $\Gamma_2$), where
$\vec{\phi}=\{x_1,x_2,x_3,x_4~|~x_1,x_2,x_3,x_4~|~x_1,x_2,x_3,x_4\}$
since $\gamma=1$ for these both irreps (see Eq.~(\ref{eq402}))
\footnote{Here $\gamma$ is the one-dimensional matrix associated
with the generator $\hat p$ (see Table~\ref{table10}).}. From the
equation $\hat\i\vec{\phi}=\vec{\phi}$, we obtain
\begin{equation}
\vec{\phi}[\Gamma_1]=\{x_1,x_2,-x_2,-x_1~|~x_1,x_2,-x_2,-x_1~|~
x_1,x_2,-x_2,-x_1\}.
 \label{eq420}
\end{equation}
This result means that the irrep $\Gamma_1$ is contained twice in
the decomposition of the mechanical representation of the considered
chain. Analogously, for the case of the irrep $\Gamma_2$, we obtain
from the equation $\hat\i\vec{\phi}=-\vec{\phi}$:
\begin{equation}
\vec{\phi}[\Gamma_2]=\{x_1,x_2,x_2,x_1~|~x_1,x_2,x_2,x_1~|~
x_1,x_2,x_2,x_1\}.
 \label{eq421}
\end{equation}
This vector also determines the two-dimensional subspace and,
therefore, the irrep $\Gamma_2$ of the group $D_3$ is contained
twice in the decomposition of the mechanical representation of our
chain.

Now let us consider the basis vectors of the two-dimensional irrep
$\Gamma_3$ from Table\,\ref{table10}. The following relations must
hold (for the complex form of this irrep)
\begin{subequations}
\label{eq440}
\begin{eqnarray}
&&\hat p\vec{\phi}=\mu\vec{\phi},\label{eq440a}\\
&&\hat p\vec{\psi}=\bar{\mu}\vec{\psi},\label{eq440b}\\
&&\hat\i\vec{\phi}=\vec{\psi},\label{eq440c}\\
&&\hat\i\vec{\psi}=\vec{\phi},\label{eq440d}
\end{eqnarray}
\end{subequations}
 where $\vec{\phi}$ and $\vec{\psi}$ are two basis
vectors of the irrep $\Gamma_3$. Comparing (\ref{eq440a}) and
(\ref{eq440b}) with (\ref{eq402}) and letting $\gamma=\mu$ and
$\gamma=\bar{\mu}$, respectively, we obtain:
\begin{equation}
\vec{\phi}=\{x_1,x_2,x_3,x_4~|~\bar{\mu}x_1,\bar{\mu}x_2,
\bar{\mu}x_3,\bar{\mu}x_4~|~\bar{\mu}^2x_1,\bar{\mu}^2x_2,
\bar{\mu}^2x_3,\bar{\mu}^2x_4\}
 \label{eq5420}
\end{equation}
and
\begin{equation}
\vec{\psi}=\{y_1,y_2,y_3,y_4~|~\mu y_1,\mu y_2, \mu y_3,\mu
y_4~|~\mu^2y_1,\mu^2y_2,\mu^2y_3,\mu^2y_4\}.
 \label{eq5421}
\end{equation}

On the other hand, the both equations (\ref{eq440c}) and
(\ref{eq440d}) lead to the same relation
$\hat\i\vec{\phi}=\vec{\psi}$. Then, from Eqs.~(\ref{eq5420}),
(\ref{eq5421}), we obtain $y_1=-\mu x_4$, $y_2=-\mu x_3$, $y_3=-\mu
x_2$, $y_4=-\mu x_1$. Therefore, the final forms of the basis
vectors of the irrep $\Gamma_3$ are:
\begin{equation}
\vec{\phi}[\Gamma_3]=\{x_1,x_2,x_3,x_4~|~\bar{\mu}x_1,\bar{\mu}x_2,
\bar{\mu}x_3,\bar{\mu}x_4~|~\bar{\mu}^2x_1,\bar{\mu}^2x_2,
\bar{\mu}^2x_3,\bar{\mu}^2x_4\},
 \label{eq430}
\end{equation}
\begin{equation}
\vec{\psi}[\Gamma_3]=\{-\mu x_4,-\mu x_3,-\mu x_2,-\mu x_1~|~
-\mu^2x_4,-\mu^2x_3,-\mu^2x_2,-\mu^2x_1,~|~-x_4,-x_3,-x_2,-x_1\}.
 \label{eq431}
\end{equation}
Each of these vectors depends on $4$ arbitrary parameters ($x_1$,
$x_2$, $x_3$, $x_4$) and, therefore, one can construct four
independent pairs of the basis vectors --- $\vec{\phi}_j[\Gamma_3]$,
$\vec{\psi}_j[\Gamma_3]$ ($j=1,2,3,4$) --- of the irrep $\Gamma_3$
\footnote{It can be done by letting $[x_1=1,x_2=0,x_3=0,x_4=0]$,
$[x_1=0,x_2=1,x_3=0,x_4=0]$, etc.}. In turn, this means that the
irrep $\Gamma_3$ is contained four times in the decomposition of the
mechanical representation of the chain with $N=12$ particles.

Taking into account the above results (see (\ref{eq420}),
(\ref{eq421}), (\ref{eq430}), (\ref{eq431})), we can conclude that
in the case $N=12$, the linearized system
$\ddot{\vec{\delta}}=\mathrm{J}(t)\cdot\vec{\delta}$ for studying
the stability of the bush B$[\hat a^4,\hat\i]$ splits into two
$2\times2$ and two $4\times4$ independent systems of differential
equations of the second order. In the real form, these systems can
be written as follows:
\begin{equation}
\begin{array}{l}
 \left\{
 \begin{array}{l}
\ddot{\nu}_1+K_3(t)\nu_1=K_1(t)\nu_2,\\
\ddot{\nu}_2+K_4(t)\nu_2=K_1(t)\nu_1,
 \end{array}
 \right.\\ \\
  \left\{
 \begin{array}{l}
\ddot{\nu}_3+K_1(t)\nu_3=K_1(t)\nu_4,\\
\ddot{\nu}_4+K_1(t)\nu_4=K_1(t)\nu_3.
 \end{array}
 \right.\\ \\
   \left\{
 \begin{array}{l}
\ddot{\nu}_5+K_6(t)\nu_5=K_1(t)\nu_6+K_2(t)\nu_7,\\
\ddot{\nu}_6+K_3(t)\nu_6=K_1(t)\nu_5,\\
\ddot{\nu}_7+K_5(t)\nu_7=K_1(t)\nu_8+K_2(t)\nu_5,\\
\ddot{\nu}_8+K_1(t)\nu_8=K_1(t)\nu_7,
 \end{array}
 \right.\\ \\
 \left\{
 \begin{array}{l}
\ddot{\nu}_9+K_6(t)\nu_9=K_1(t)\nu_{10}+K_2(t)\nu_{11},\\
\ddot{\nu}_{10}+K_3(t)\nu_{10}=K_1(t)\nu_9,\\
\ddot{\nu}_{11}+K_5(t)\nu_{11}=K_1(t)\nu_{12}+K_2(t)\nu_9,\\
\ddot{\nu}_{12}+K_1(t)\nu_{12}=K_1(t)\nu_{11},
 \end{array}
 \right.
\end{array}
 \label{eq1234}
\end{equation}
where
\[
\begin{array}{l}
K_1(t)=1-2A(t)+2B(t),\\
K_2(t)=\sqrt{3}[-\frac{1}{2}+2A(t)],\\
K_3(t)=3+6A(t)+2B(t),\\
K_4(t)=3-2A(t)-6B(t),\\
K_5(t)=\frac{5}{2}-2A(t)-4B(t),\\
K_6(t)=\frac{3}{2}-2A(t).
\end{array}
\]

The generalization to the case of an arbitrary $N$ (note that
$N\mod4=0$ must hold!) is trivial: each one-dimensional irrep is
contained twice in the decomposition of the mechanical
representation of the $N$-particle chain, while each two-dimensional
irrep \footnote{The number of two-dimensional irreps of the bush
symmetry group $[\hat a^4,\hat\i]=D_{\frac{N}{4}}$ increases with
increasing $N$.} is contained four times in it.

\begin{figure}
\includegraphics[width=100mm,height=100mm]{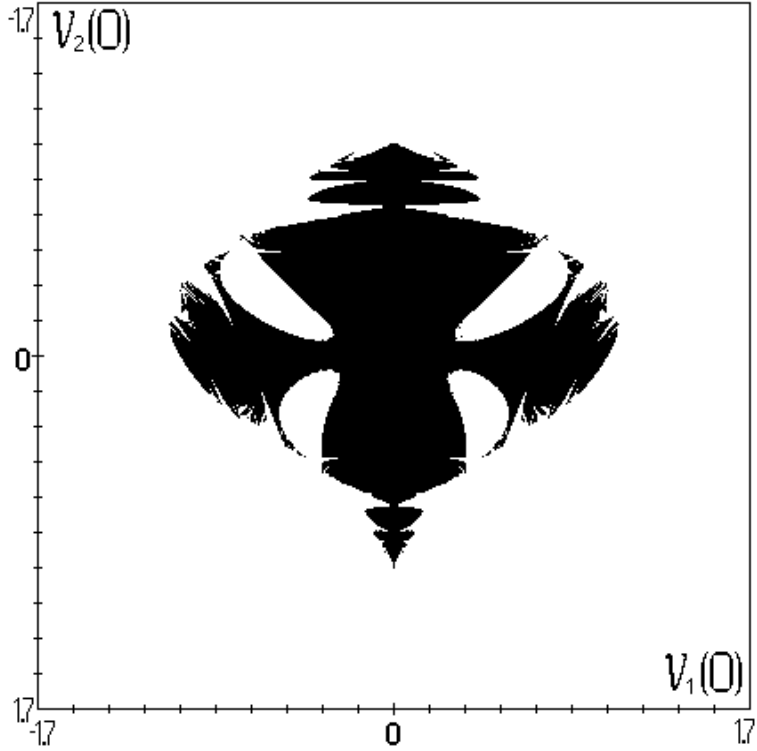}
\caption{\label{fig2} Stability diagram for the bush B$[a^4,i]$ in
the FPU-$\alpha$ chain.}
\end{figure}

In conclusion, in Fig.\,\ref{fig2} we reproduce the stability
diagram for the two-dimensional bush B$[\hat a^4,\hat\i]$ from our
paper \cite{FPU2}. This diagram corresponds to the FPU-$\alpha$
chain with $N=12$ particles. It represents a planar section of the
four-dimensional stability domain in the space of the initial
conditions $\nu_1(0)$, $\nu_2(0)$, $\dot{\nu}_1(0)$,
$\dot{\nu}_2(0)$, where $\nu_1(t)$ and $\nu_2(t)$ are two modes of
the considered bush.

In producing Fig.\,\ref{fig2}, we specify $\dot{\nu}_1(0)=0$,
$\dot{\nu}_2(0)=0$, and change $\nu_1(0)$, $\nu_2(0)$ in some
interval near their zero values. The stability domain, resembling a
beetle, is drown in \emph{black} color in the plain $\nu_1(0)$,
$\nu_2(0)$. The bush B$[\hat a^4,\hat\i]$ losses its stability (and
transforms into another bush of higher dimension), when we cross the
boundary of the black region in any direction. From Fig.\,\ref{fig2}
it is obvious, how nontrivial stability domain for a bush of modes
can be.

A detailed description of the stability domains for one-dimensional
and two-dimensional bushes of modes in both FPU-$\alpha$ and
FPU-$\beta$ chains can be found in \cite{FPU2}.

\section{Conclusion \label{Concl}}

All the exact dynamical regimes in $N$-particle mechanical system
with discrete symmetry can be classified by the \emph{subgroups}
$G_j$ of the parent group $G_0$, i.e. the symmetry group of its
equations of motion. Actually, each subgroup $G_j$ singles out a
certain \emph{invariant manifold} which, being decomposed into the
basis vectors of the irreducible representations of the group $G_0$,
is termed as a ``bush of modes'' \cite{DAN1,DAN2,PhysD}.

The bush B$[G_j]$, representing an $n$-dimensional vibrational
regime, can be considered as a dynamical object characterized by its
displacement pattern of all the particles from their equilibrium
positions, by the appropriate dynamical equations and the domain of
the stability. One-dimensional bushes are symmetry-determined
similar nonlinear normal modes introduced by Rosenberg \cite{Ros}
(see also \cite{IntJ}). For Hamiltonian systems, the energy of the
initial excitation turns out to be ``trapped'' in the bush and this
is a phenomenon of energy localization in the modal space.

The different aspects of the bush theory were developed in
\cite{DAN1,DAN2,PhysD,IntJ,ENOC,C60,Octa}. Bushes of vibrational
modes (invariant manifolds) in the FPU chains were discussed in
\cite{FPU1,FPU2,BR,PR,Shin}.

The stability analysis of a given bush B$[G]$ reduces to studying
the linearized (in the vicinity of the bush) dynamical equations
$\ddot{\vec{\delta}}=\mathrm{J}(t)\cdot\vec{\delta}$. In the present
paper, we prove (Theorem\,\ref{theor1}) that the symmetry group of
the linearized system
$\ddot{\vec{\delta}}=\mathrm{J}(t)\cdot\vec{\delta}$ turns out to be
precisely the symmetry group $G$ of the considered bush B$[G]$. This
result allows one to apply the well-known Wigner theorem about the
specific structure of the matrix ($\mathrm{J}(t)$, in our case)
commuting with all the matrices of a fixed representation
(mechanical representation, in our case) of a given group. According
to the above theorem one can split effectively the linearized system
$\ddot{\vec{\delta}}=\mathrm{J}(t)\cdot\vec{\delta}$ into a number
of independent subsystems of differential equations with
time-dependent coefficients.

We want to emphasize that this symmetry-related method for splitting
the linearized systems arising in the linear stability analysis of
the dynamical regimes is suitable for \emph{arbitrary} nonlinear
mechanical systems with discrete symmetry. Such a decomposition
(splitting) of the linearized system
$\ddot{\vec{\delta}}=\mathrm{J}(t)\cdot\vec{\delta}$ is especially
important for the \emph{multidimensional} bushes of modes,
describing \emph{quasiperiodic} vibrational regimes, which cannot be
treated with the aid of the Floquet method. Indeed, in this case, we
need to integrate the differential equations with time-dependent
coefficients over large time interval, unlike the case of periodic
regimes where we can solve the appropriate differential equations
over only one period to construct the monodromy matrix.

The above method is applied for studying the stability of some
dynamical regimes (bushes of modes) in the monoatomic chains. For
this specific mechanical systems, we prove Theorem\,\ref{theor2}
which allows one to find very simply the upper bound of dimensions
of the independent subsystems obtained after splitting the
linearized system
$\ddot{\vec{\delta}}=\mathrm{J}(t)\cdot\vec{\delta}$. Indeed,
according to this theorem, the dimension of each such subsystem does
not exceed the integer $m$ determining the ratio of the volumes of
the primitive cell of the chain in the vibrational state,
corresponding to the given bush B$[G]=$B$[\hat a^m,\dots]$, and the
equilibrium state.

Taking into account any other symmetry elements of the considered
bush allows to reduce the dimensions of at least some of the above
discussed subsystems. We illustrate this fact comparing the
stability analysis of the $\pi$-mode (zone boundary mode) for the
FPU-$\alpha$ and FPU-$\beta$ chains.

\begin{acknowledgments}
We are very grateful to Prof.\ V.P.~Sakhnenko for useful discussions
and for his friendly support, and to O.E.~Evnin for his valuable
help with the language corrections in the text of this paper.
\end{acknowledgments}

\end{document}